\documentclass[a4paper,11pt]{article}
\usepackage{amsmath,amssymb,amsthm,amsxtra,overpic,bbm,bbold,bm,epsfig,multirow}
\usepackage{color,ulem,graphicx, caption, subcaption,hyperref,cite,float}

\allowdisplaybreaks

\begin{document}
\begin{titlepage}
\begin{center}
{\bf\Large
\boldmath{
Revisiting Trimaximal TM2 mixing in two $A_4$ modular symmetries}
}
\\[12mm]
Han~Zhang$^{1,2,3}$\footnote{E-mail: \texttt{zhanghan216@mails.ucas.ac.cn}}
and
Ye-Ling~Zhou$^{1,}$\footnote{E-mail: \texttt{zhouyeling@ucas.ac.cn}}
\\[-2mm]
\end{center}
\vspace*{0.50cm}
\centerline{$^{1}$\it School of Fundamental Physics and Mathematical Sciences,}
\centerline{\it Hangzhou Institute for Advanced
Study, UCAS, Hangzhou 310024, China}
\centerline{$^{2}$\it Institute of Theoretical Physics, Chinese Academy of Sciences, Beijing 100190, China}
\centerline{$^{3}$\it University of Chinese Academy of Sciences, Beijing 100049, China}

\vspace*{1.20cm}

\begin{abstract}
{\noindent
We discuss a minimal lepton flavour model with two modular $A_4$ symmetries, one acting on neutrinos and the other acting on charged leptons. 
Two corresponding moduli fields are restricted at stabilisers. To avoid vanishing or degenerate lepton masses, the stabiliser in the charged lepton sector always preserves a $Z_3$ symmetry, and that in the neutrino preserves a $Z_2$ symmetry. By scanning all viable stabilisers, a unique Trimaximal TM2 mixing is predicted. However, the prediction of the lightest neutrino mass and the mass parameter in neutrinoless double beta decay, as well as two Majorana phases, are different and classified into three cases.
}
\end{abstract}
\end{titlepage}

\section{Introduction}

The origin of lepton flavour mixing is still a main puzzle in neutrino physics. 
For decades, non-Abelian discrete symmetries have been regarded as the most convincing way to understand this puzzle. 
In a traditional framework of flavour model construction, flavons for charged lepton and neutrino masses are introduced separately \cite{Altarelli:2005yp,Altarelli:2005yx}. They gain different vacuum expectation values (VEVs), break the flavour symmetry along different directions and give rise to the lepton flavour mixing.
The unbroken residual symmetries can naturally predict some special flavour patterns \cite{Ge:2011ih,Ge:2011qn}. In particular, Trimaximal mixing 1 (TM1) \cite{Albright:2008rp} and Trimaximal mixing 2 (TM2) \cite{Grimus:2008tt}, which preserve the first and second columns of the Tri-bimaximal mixing \cite{Harrison:2002er,Xing:2002sw}, respectively, are still consistent with current oscillation data. 

The modular invariance approach, where a finite modular symmetry $\Gamma_N$ plays the role of the flavour symmetry, provides an alternative way to solve the lepton flavour puzzle \cite{Feruglio:2017spp}. The main feature is that the Yukawa couplings and fermion mass matrices arise from modular forms which depend on the value of the modulus field. In addition, both the couplings and fields transform under a finite modular group $\Gamma_N$. These modular forms transforms as multiplets of $\Gamma_N$, replacing the role of flavons in the traditional approach. 
Along this approach, successful lepton flavour models have initiated in 
$\Gamma_2\simeq S_3$ \cite{Kobayashi:2018vbk}, 
$\Gamma_3\simeq A_4$ \cite{Criado:2018thu}, 
$\Gamma_4\simeq S_4$ \cite{Penedo:2018nmg}, 
$\Gamma_5 \simeq A_5$ \cite{Novichkov:2018nkm} and 
$\Gamma_7 \simeq PSL(2, Z_7)$ \cite{Ding:2020msi}. See also in \cite{Kobayashi:2018wkl,Kobayashi:2019rzp,Okada:2019xqk,
Kobayashi:2018scp,deAnda:2018ecu,Okada:2018yrn,Novichkov:2018yse,Nomura:2019jxj,Okada:2019uoy,Nomura:2019yft,Ding:2019zxk,Okada:2019mjf,Nomura:2019lnr,Kobayashi:2019xvz,Asaka:2019vev,Ding:2019gof,Zhang:2019ngf,Nomura:2019xsb,Wang:2019xbo,Kobayashi:2019gtp,King:2020qaj,Ding:2020yen,Okada:2020rjb,Nomura:2020opk,Asaka:2020tmo,Okada:2020brs,Yao:2020qyy,Feruglio:2021dte,
Novichkov:2018ovf,deMedeirosVarzielas:2019cyj,Kobayashi:2019mna,Criado:2019tzk,King:2019vhv,Wang:2019ovr,Wang:2020dbp,Qu:2021jdy,
Ding:2019xna} for examples of successful models and a recent review in \cite{Ding:2023htn}.

Formalism of the modular invariance approach was generalised to multiple modular symmetries in \cite{deMedeirosVarzielas:2019cyj,King:2019vhv}, with earlier phenomenological studies considered in \cite{Novichkov:2018ovf,Novichkov:2018yse}. 
By introducing bi-triplet scalars as bridges to connect different modular groups, multiple modular symmetries can be spontaneously broken and at the lower energy scale, the Lagrangian appears as an effective theory in a single modular group of multiple moduli fields \cite{deMedeirosVarzielas:2019cyj}.
As more than moduli fields are allowed, moduli fields in the charged lepton sector and neutrino sector can be different and gain different VEVs. They break the modular symmetry in the charged lepton and neutrino mass matrices in different ways.

A modulus field may gain the VEV at a stabiliser. The latter by definition is a fixed point invariant under part of the modular transformations. The lepton mass matrix at the stabiliser can preserve a residual symmetry which is a subset of the modular symmetry. Implications of residual modular symmetries were proposed to derive simple structures of the modular form \cite{Novichkov:2018ovf}. By assuming different residual symmetries, i.e., $Z_3$ and $Z_2$ in the charged lepton and neutrino sectors, respectively, TM2 mixing was discussed in the framework of $A_4$ \cite{Novichkov:2018yse}. Concrete models to derive TM1 mixing were achieved in two $S_4$'s \cite{King:2019vhv} and TM2 mixing in two $A_4$'s \cite{deMedeirosVarzielas:2021pug}. The approach of multiple modular symmetries was extended to quark flavours \cite{Kikuchi:2023jap,Petcov:2023vws,deMedeirosVarzielas:2023crv}. 
Those in the GUT framework were studied in \cite{King:2021fhl,deMedeirosVarzielas:2023ujt}. 
Modulus stabilisation mechanism was investigated in the multiple moduli framework \cite{King:2023snq}.

In this paper, we will give a comprehensive study on lepton flavour mixing in two modular $A_4$ symmetries with moduli fields fixed at any stabilisers. The rest of the paper is organised as follows. We briefly review the framework of multiple modular symmetries in Section~\ref{sec:2}. In Section~\ref{sec:3} we present an economical model in $A_4^l \times A_4^\nu$ symmetries. In Section~\ref{sec:4} we give more details when moduli fields are fixed at stabilisers and give both analytical and numerical results. We give the summary in Section~\ref{sec:5}.


\section{Framework of modular $A_4$ flavour models} \label{sec:2}

A modular transformation $\gamma$ is defined as a linear fractional transformation acting on the modulus space $\tau = {\rm Re}\, \tau + i\, {\rm Im}\, \tau$ with ${\rm Im}\, \tau >0$, 
\begin{eqnarray} \label{eq:modular_transformation}
\gamma:~ 
\tau \to \gamma \tau = \frac{a \tau + b}{c \tau + d}\,,
\end{eqnarray}
where $a$, $b$, $c$, and $d$ are integers satisfying $ad-bc=1$. All such transformations form the modular group $\overline{\Gamma}$, which is a discrete and infinite group. 
The group has two generators, $S$ and $T$. They act on the modulus $\tau$ in the following way, 
\begin{eqnarray}
S: \tau \to -\frac{1}{\tau} \,, \quad
T: \tau \to \tau + 1\,,
\end{eqnarray}
respectively. It is straightforward to check they satisfies $S^2 = (S T)^3 = \mathbf{1}$. 
Respect to the integer coefficients $a$, $b$, $c$, and $d$, each element $\gamma$ of $\overline{\Gamma}$ can be represented by a $2\times2$ matrix with entries, saying, 
\begin{eqnarray}
\overline{\Gamma} = \left\{ \begin{pmatrix} a & b \\ c & d \end{pmatrix} / (\pm \mathbf{1})\,,~ a, b, c, d \in \mathbb{Z}, ~~  ad-bc=1  \right\} \,.
\end{eqnarray}
In particular, the generators, are represented by $2\times 2$ matrices as follows
\begin{eqnarray}
S_\tau=\eta \begin{pmatrix} 0 & 1 \\ -1 & 0 \end{pmatrix}\,, \hspace{1cm}
T_\tau=\eta \begin{pmatrix} 1 & 1 \\ 0 & 1 \end{pmatrix} \,,
\end{eqnarray}
regardless to the sign difference $\eta = \pm 1$. 
The group $A_4$ is a finite subgroup of $\overline{\Gamma}$. It is obtained by requiring $T^3 = \mathbf{1}$. It is isotropic to the quotient group, $A_4 \simeq \Gamma_3 \equiv \overline{\Gamma} / \overline{\Gamma}(3)$, where $\overline{\Gamma}(3)$ is an infinite subgroup of $\overline{\Gamma}$ labelled as 
\begin{eqnarray}
\overline{\Gamma}(3) = \left\{ \begin{pmatrix} a & b \\ c & d \end{pmatrix} \in \overline{\Gamma}, ~~  \begin{pmatrix} a & b \\ c & d \end{pmatrix} = \begin{pmatrix} 1 & 0 \\ 0 & 1 \end{pmatrix} ~~ ({\rm mod}~ 3) \right\} \,.
\end{eqnarray}
Using this correlation, we can write any element $\gamma$ of $A_4$ in the form
\begin{eqnarray} \label{eq:rep_1}
\eta \begin{pmatrix} 3 k_a +a & 3 k_b +b \\ 3 k_c +c & 3 k_d +d \end{pmatrix} \,,
\end{eqnarray} 
regardless to $\eta$ and any integers $k_a$, $k_b$, $k_c$ and $k_d$ which satisfy $3 k_a k_d + a k_d + d k_a = 3 k_b k_c + b k_c + c k_b$ and $\eta= \pm 1$. Here, the sign difference $\eta$ and integers are mathematical redundancies induced by the representation. Changing their values does not lead to any physical difference. 
 
We follow the widely-considered framework of $\mathcal{N}=1$ supersymmetry with the $A_4$ modular symmetries, where the action takes the form
\begin{eqnarray}
S = \int d^4 d^2 \theta d^2 \bar{\theta}\, K (\phi_i, \bar{\phi}_i;\tau, \bar{\tau}) + \int d^4 x d^2 \theta \, W (\phi_i; \tau) + {\rm h.c.}.
\end{eqnarray}
Here, $K (\phi_i, \bar{\phi}_i;\tau,\bar{\tau})$ is the K\"ahler potential. We assume the minimal form 
\begin{eqnarray}
K (\phi_i, \bar{\phi}_i;\tau, \bar{\tau}) = - h \log \det (-i \tau + i \bar{\tau}) + \sum_i \frac{\phi_i \bar{\phi}_i}{(-i\tau+ i\bar{\tau})^{2k_i}} \,,
\end{eqnarray} 
which can leave the kinetic terms of the supermultiplets $\phi_i$ and the modulus field as
\begin{eqnarray}
h\, \frac{\partial_\mu \bar{\tau} \partial^\mu \tau}{\langle -i\tau+ i \bar{\tau} \rangle^2 } 
+ \sum_i 
\frac{\partial_\mu \bar{\phi}_i \partial^\mu \phi_i}{\langle -i\tau+ i \bar{\tau} \rangle^2 } 
\end{eqnarray}
with $2k_i$ the modular weight of $\phi_i$ and $h>0$ a coefficient of two units of mass dimension. Taking more complicated form of the K\"ahler potential introduces additional free parameters and as a consequence renders the model less predictive \cite{Chen:2019ewa}. 
In the case of a single modular $A_4$ symmetry, the superpotential $W(\phi_i;\tau)$ is a function of the modulus field $\tau$ and superfields $\phi_i$, which is invariant under the modular transformation \cite{Ferrara:1989bc}. 
The superpotential can be expanded in powers of the superfields $\phi_i$, i.e.,
\begin{eqnarray}
W(\phi_i;\tau) = \sum_n \sum_{\{i_1, \cdots, i_n\}} \sum_{I_Y} \left( Y_{I_Y} \phi_{i_1} \cdots \phi_{i_n} \right)_{\mathbf{1}} \,,
\end{eqnarray}
where $Y_{I_Y}$ represents a collection of coefficients of the couplings. The chiral superfield $\phi_i$, as a function of $\tau$ transforms as \cite{Ferrara:1989bc},
\begin{eqnarray}
\gamma:  \phi_i(\tau) \to \phi_i(\gamma\tau) = (c\tau + d)^{-2k_i} \rho_{I_i}(\gamma) \phi_i(\tau)\,,
 \label{eq:field_transformation}
\end{eqnarray}
where $-2k_i$ is the modular weight of $\phi_i$, ${I_i}$ denotes the representation of $\phi_i$ and $\rho_{I_i}(\gamma)$ is a unitary representation matrix of $\gamma$ with $\gamma \in A_4$ depending on the representation of $\phi_i$ assigned in $A_4$. 
The coefficients $Y_{I_Y}$ do not have to be constants, but modular forms in the framework of modular symmetries. 
A modular form of weight $2k$ and level $3$ is a holograph function of the modulus $\tau$. These modular forms form multiplets in the quotient group $A_4 \simeq \overline{\Gamma}/\overline{\Gamma}(3)$. For a modular transformation $\gamma \in A_4$, the transformation appears to be 
\begin{eqnarray}
\gamma: Y^{(2k)}(\tau) \to Y^{(2k)}(\gamma \tau) = (c\tau + d)^{2k} \rho_{I_Y}(\gamma) Y^{(2k)}(\tau) \,,
\end{eqnarray}
where $\rho_{I_Y}(\gamma)$ is the unitary matrix of $\gamma$ in the representation $I_Y$. 
At $2k=2$, there are three modular forms which form an irreducible triplet of $A_4$ and can be expressed in terms of the Dedekind eta functions and its derivative (see Appendix~\ref{sec:modular-forms}). We denoted them as $Y = (Y_1, Y_2,Y_3)^{\rm T}$, where $Y_2^2 + 2 Y_1 Y_3 = 0$ is satisfied. At $2k=4$, there are 5 modular forms, forming a trivial singlet $\bf 1$, a non-trivial singlet $\bf 1'$ and a triplet $\bf 3$ of $A_4$. We denote them as follows
\begin{eqnarray}
    X &=& Y_1^2 + 2 Y_2 Y_3 \,, \nonumber\\
    X' &=& Y_3^2 + 2 Y_1 Y_2\,, \nonumber\\
    Z &=& (Z_1, Z_2, Z_3)^{\rm T} \equiv  (Y_1^2-Y_2 Y_3, Y_3^2-Y_1 Y_2, Y_2^2-Y_1 Y_3)^{\rm T} \,,
\end{eqnarray}
where the symbol $^{\rm T}$ represents the transpose of a vector or matrix.
Note the singlet $\bf 1''$ vanishes since $Y_2^2 + 2 Y_1 Y_3 = 0$.
To be invariant under the modular transformation, the Yukawa couplings $Y_{I_Y}$ can be arranged as modular forms of modular weight $2k_Y = 2k_{i_1} + \cdots + 2 k_{i_n}$ satisfied, and can be reduced to a set of irreducible representations of $A_4$. 



We will discuss flavour models with two modular symmetries $\overline{\Gamma}^{l}$ and $\overline{\Gamma}^{\nu}$. Their generators ($S$, $T$) are denoted by ($S_l$, $T_l$) and ($S_\nu$, $T_\nu$), respectively. We only consider the direct product of them, and the moduli fields are denoted as $\tau_l$ and $\tau_\nu$, respectively. Any two modular transformations $\gamma_l \times \gamma_\nu$ in $\overline{\Gamma}^l \times \overline{\Gamma}^\nu$ take forms as 
\begin{eqnarray} 
&&\gamma_l \times \gamma_\nu: (\tau_l,\tau_\nu) \to (\gamma_l \tau_l,\gamma_\nu \tau_\nu) = 
\left( \frac{a_l \tau_l + b_l}{c_l \tau_l + d_l}, \frac{a_\nu \tau_\nu + b_\nu}{c_\nu \tau_\nu + d_\nu} \right) \,.
\end{eqnarray}
Two finite modular groups $A_4^l$ and $A_4^\nu$ can be obtained by imposing $T_l^3 = T_\nu^3 = \mathbf{1}$. 
The K\"ahler potential $K (\phi_i, \bar{\phi}_i;\tau_l,\bar{\tau}_l, \tau_\nu,\bar{\tau}_\nu)$ and superpotential $W(\phi_i;\tau_l, \tau_\nu)$ are now expanded as 
\begin{eqnarray}
&&K (\phi_i, \bar{\phi}_i;\tau_l,\bar{\tau}_l, \tau_\nu,\bar{\tau}_\nu) =  - h_l \log \det (-i \tau_l + i \bar{\tau}_l) - h_\nu \log \det (-i \tau_\nu + i \bar{\tau}_\nu)  \nonumber\\ 
&&\hspace{4cm}+ \sum_i \frac{\phi_i \bar{\phi}_i}{(-i\tau_l+ i\bar{\tau}_l)^{2k_{l,i}} (-i\tau_\nu+ i\bar{\tau}_\nu)^{2k_{\nu,i}}} \nonumber\\
&&W(\phi_i;\tau_l, \tau_\nu) = \sum_n  \sum_{\{i_I, \cdots, i_n\}} 
\left(Y_{(I_{Y,l}, I_{Y,\nu})} \phi_{i_1} \cdots \phi_{i_n} \right)_{(\mathbf{1}, \mathbf{1})} \,,
\end{eqnarray}
respectively, where $2k_{l,i}$ and $2k_{\nu,i}$ are modular weights of $\phi_i$ in $\overline{\Gamma}^l \times \overline{\Gamma}^\nu$, respectively.
The chiral field $\phi_i$ and
the modular form $Y_{(I_{Y,l}, I_{Y,\nu})}$ respectively transform as 
\begin{eqnarray} \label{eq:field_form_transformation}
 \phi_i(\tau_l, \tau_\nu) &\to& \phi_i(\gamma_l\tau_l, \gamma_\nu \tau_\nu) 
\nonumber\\
&& = (c_l\tau_l + d_l)^{-2k_{i,l}}  (c_\nu\tau_\nu + d_\nu)^{-2k_{i,\nu}} \rho_{I_{i,l}}(\gamma_l) \phi_i(\tau_l, \tau_\nu) \rho^{\rm T}_{I_{i,\nu}}(\gamma_\nu)\,, \nonumber\\
\hspace{-5mm}
Y_{(I_{Y,l}, I_{Y,\nu})}(\tau_l, \tau_\nu) &\to& Y_{(I_{Y,l}, I_{Y,\nu})}(\gamma_l \tau_l, \gamma_\nu \tau_\nu) \nonumber\\
&&= (c_l\tau_l + d_l)^{2k_{Y,l}} (c_\nu \tau_\nu + d_\nu)^{2k_{Y,\nu}}
 \rho_{I_{Y,l}}(\gamma_l) Y_{(I_{Y,l}, I_{Y,\nu})}(\tau_l, \tau_\nu) \rho^{\rm T}_{I_{Y,\nu}}(\gamma_\nu) \,.
\end{eqnarray}
Here, we have arranged $\phi_i$ and $Y_{(I_{Y,l}, I_{Y,\nu})}$ as matrices instead of vectors, and let $\gamma_l$ act on them vertically and $\gamma_\nu$ act on them horizontally. 
Including twin modular symmetries allows us to break modular symmetries into different subgroups for charged lepton sector and neutrino sector respectively. 

\section{Flavour Model} \label{sec:3}

The flavour model is outlined below. It is invariant under two modular symmetries, $A_4^l$ and $A_4^\nu$, with moduli fields labelled by $\tau_l$ and $\tau_\nu$, respectively. 
We follow the sketch as introduced in \cite{King:2019vhv,deMedeirosVarzielas:2019cyj} that the multiple modular symmetries are first broken to a single modular symmetry and later is then further broken after the moduli fields gain different VEVs. The lepton flavour mixing arises due to the misalignment of the VEVs.

Lepton chiral superfields are arranged in the left panel of Table~\ref{tab:particle_content}. They are summarised below: 1) the right-handed charged leptons $e^c$, $\mu^c$ and $\tau^c$ are singlets $\mathbf{1}$, $\mathbf{1}''$ and $\mathbf{1}'$ of $A_4^l$, and trivial singlets $\mathbf{1}$ of $A_4^\nu$, and have weights $2k_l=+2$ or $+4$ for $A_4^l$, $2k_\nu=+2$ for $A_4^\nu$; 2) the three lepton doublets form a triplet $L=(L_1, L_2, L_3)^{\rm T}$ of $A_4^l$ with zero weight, and a trivial singlet $\mathbf{1}$ of $A_4^\nu$ with weight $2k_\nu=-2$; 3) We introduce three right-handed neutrinos $\nu^c$ which form a triplet $(\nu^c_1, \nu^c_2, \nu^c_3)^{\rm T}$ of $A_4^\nu$ with weight $2k_\nu=+2$. 
In addition, we introduce a bi-triplet scalar $\Phi \sim (\mathbf{3}, \mathbf{3})$, which are crucial to achieve the breaking $A_4^l \times A_4^\nu \to A_4$. 
The modular-invariant superpotential terms to generate lepton masses are given by
\begin{eqnarray} \label{eq:superpotential_lepton}
W &\supset& \left[ y_e\, e^c (Y_l(\tau_l) L)_{\bf 1} + y_\mu\, \mu^c (Y_l(\tau_l) L)_{\bf 1'} + y_\tau\, \tau^c (Y_l(\tau_l) L)_{\bf 1''} \right] H_d \nonumber\\
&+& \frac{y_\nu}{\Lambda} (\nu^c \Phi L)_{({\bf 1,1})}  H_u 
+ \frac{1}{2} M_{\mathbf{1}}(\tau_\nu) (\nu^c \nu^c)_{\mathbf{1}} + \frac{1}{2} M_{\mathbf{1}'}(\tau_\nu) (\nu^c \nu^c)_{\mathbf{1}''} + \frac{1}{2} M_{\mathbf{3}}(\tau_\nu) (\nu^c \nu^c)_{\mathbf{3}}\,,
\end{eqnarray}
where $y_{e,\mu,\tau}$ are free parameters which can always be chosen to be real and positive by rotating phases of $e^c$, $\mu^c$ and $\tau^c$. Yukawa couplings and right-handed neutrino masses are arranged as modular forms to be consistent with the invariance of modular symmetries, which are presented in the right panel of Table~\ref{tab:particle_content}. $Y_l$ is a $\mathbf{3}$-plet modular form of $A_4^l$. $y_\nu$ can only be a modulus-independent coefficient in this model instead of a modular form. $M_{\mathbf{1}}(\tau_\nu) = \mu_1 X$, $M_{\mathbf{1}'}(\tau_\nu) = \mu_{1'} X'$ and $M_{\mathbf{3}}(\tau_\nu) = \mu_3 Z$ represents $\mathbf{1}$-, $\mathbf{1}'$- and $\mathbf{3}$-plets modular forms appearing in right-handed neutrino mass terms and $\mu_1$, $\mu_{1'}$ and $\mu_3$ are free parameters of a mass dimension.

\begin{table}[t!] 
\begin{center}
{\begin{tabular}{| l | c c c c|}
\hline \hline
Fields & $A_4^l$ & $A_4^\nu$ & $2k_l$ & $2k_\nu$\\ 
\hline \hline
$e^c$ & $\mathbf{1}$ & $\mathbf{1}$ & $+2(+4)$ & $+2$ \\
$\mu^c$ & $\mathbf{1}''$ & $\mathbf{1}$ & $+2(+4)$ & $+2$ \\
$\tau^c$ & $\mathbf{1}'$ & $\mathbf{1}$ & $+2(+4)$ & $+2$ \\
$L$ & $\mathbf{3}$ & $\mathbf{1}$ & $0$ & $-2$\\
$\nu^c$ & $\mathbf{1}$ & $\mathbf{3}$ & $0$ & $+2$ \\
\hline 
$\Phi$ & $\mathbf{3}$ & $\mathbf{3}$ & $0$ & $0$ \\
\hline 
$H_{u,d}$ & $\mathbf{1}$ & $\mathbf{1}$ & $0$ & $0$ \\
\hline \hline
\end{tabular}
\hspace{2mm}
\begin{tabular}{| l | c c c c|}
\hline \hline
Yukawas / masses & $A_4^l$ & $A_4^\nu$ & $2k_l$ & $2k_\nu$\\
\hline \hline
$Y_l(\tau_l)$ & $\mathbf{3}$ & $\mathbf{1}$ & $+2(+4)$ & $0$ \\
$M_{\mathbf{1}}(\tau_\nu)$ & $\mathbf{1}$ & $\mathbf{1}$ & $0$ & $+4$ \\
$M_{\mathbf{1}'}(\tau_\nu)$ & $\mathbf{1}$ & $\mathbf{1}'$ & $0$ & $+4$ \\
$M_{\mathbf{3}}(\tau_\nu)$ & $\mathbf{1}$ & $\mathbf{3}$ & $0$ & $+4$ \\
\hline \hline 
\multicolumn{5}{c}{}
\end{tabular}}
\caption{Representations arrangements for fields, Yukawa couplings and Majorana masses for right-handed neutrinos.} \label{tab:particle_content}
\end{center}

\end{table}

The scalar $\Phi$ is used for the connection between two $A_4$'s and its VEV is the key to break two groups to a single group. 
The operator $(\nu^c \Phi L)_{({\bf 1,1})}H_u$ is expanded as 
\begin{eqnarray}
(\nu^c_{1}, \nu^c_{2}, \nu^c_{3}) P_{23} \begin{pmatrix}
\Phi_{11} & \Phi_{12} & \Phi_{13} \\ 
\Phi_{2 1} & \Phi_{2 2} & \Phi_{2 3} \\ 
\Phi_{3 1} & \Phi_{3 2} & \Phi_{3 3} 
\end{pmatrix} P_{23} 
\begin{pmatrix}
L_1 \\ 
L_2 \\ 
L_3 
\end{pmatrix} H_u\,.
\end{eqnarray}
In the main text, we will fix the VEV of $\Phi$ at $\langle \Phi \rangle_{\alpha i} = v_\Phi (P_{23})_{\alpha i}$, where $\alpha,i = 1,2,3$  and $P_{23}$ is a permutation matrix
\begin{eqnarray}
P_{23} = \begin{pmatrix} 1 & 0 & 0 \\ 0 & 0 & 1 \\ 0 & 1 & 0 \end{pmatrix} \,.
\end{eqnarray} 
Note that the VEV is not unique. We discuss the VEV alignment and the equivalence of all VEVs up to basis transformation in Appendix~\ref{sec:vacuum-alignments}.
Fixing $\Phi$ at its VEV, this term is left with $y_D (\nu^c_1 L_1 +  \nu^c_3 L_2 + \nu^c_2 L_3 ) H_u$, where $y_D = y_\nu v_\Phi / \Lambda$. We can rewrite it as $(\nu^c)^{\rm T} y_D P_{23} L H_u$.

Now the superpotential terms become 
\begin{eqnarray}
W|_{\langle \Phi \rangle} &\supset& \left[ e^c Y_e(\tau_l) L + \mu^c Y_\mu(\tau_l) L + \tau^c Y_\tau(\tau_l) L \right] H_d 
+  (\nu^c)^{\rm T} y_D P_{23} L H_u  
+ \frac{1}{2} (\nu^c)^{\rm T} M_{\nu^c}(\tau_\nu) \nu^c \,.
\end{eqnarray}
In the charged lepton sector, $Y_e$, $Y_\mu$ and $Y_\tau$ are rearrangements of the triplet modular form $Y_l$. In Table~\ref{tab:particle_content}, we have listed two values of modular weight for $Y_l$. For modular weight $2k_l=2$, we have $Y_e=y_e (Y_1,Y_3,Y_2)^{\rm T}$,$Y_\mu=y_\mu (Y_2, Y_1, Y_3)^{\rm T}$, and $Y_\tau = y_\tau (Y_3, Y_2, Y_1)^{\rm T}$. The mass matrix for charged leptons is given by
\begin{eqnarray} \label{eq:M_E_2}
M_l^* = 
\begin{pmatrix}
Y_1 & Y_2 & Y_3 \\
Y_3 & Y_1 & Y_2 \\
Y_2 & Y_3 & Y_1
\end{pmatrix} {\rm diag}\{y_e, y_\mu, y_\tau\} v_d \,.
\end{eqnarray}
Here and below, we use the left-right convention, (e.g., $\overline{l_L} M_l l_R + {\rm h.c.}$) and thus the complex conjugation $*$ appears in some mass matrices. 
Similarly, for modular weight $2k_l =4$, we have the Yukawa form modified to $Y_e=y_e (Z_1,Z_3,Z_2)^{\rm T}$,$Y_\mu=y_\mu (Z_2, Z_1, Z_3)^{\rm T}$, and $Y_\tau = y_\tau (Z_3, Z_2, Z_1)^{\rm T}$. Then the charged lepton mass matrix is modified to
\begin{eqnarray} \label{eq:M_E_4}
M_l^* = 
\begin{pmatrix}
Z_1 & Z_2 & Z_3 \\
Z_3 & Z_1 & Z_2 \\
Z_2 & Z_3 & Z_1
\end{pmatrix} {\rm diag}\{y_e, y_\mu, y_\tau\} v_d \,.
\end{eqnarray}
The Majorana mass matrix for right-handed neutrinos $M_R = M_{\nu^c}^*$ is in general expressed as 
\begin{eqnarray} \label{eq:M_RHN}
M_R^*  = 
\begin{pmatrix}
\mu_1 X & 0 & \mu_{1'} X' \\ 
0 & \mu_{1'} X' & \mu_1 X \\ 
\mu_{1'} X' & \mu_1 X & 0
\end{pmatrix}
+
\mu_3 \begin{pmatrix}
2 Z_1 & -Z_3 & -Z_2 \\ -Z_3 & 2Z_2 & -Z_1 \\ -Z_2 & -Z_1 & 2Z_3
\end{pmatrix}
\,. 
\end{eqnarray}
The effective neutrino mass matrix is obtained by the Type-I seesaw mechanism after the moduli fields and Higgs gains VEVs,
\begin{eqnarray}
M_\nu=-M_D M_R^{-1} M_D^{\rm T} \,,
\end{eqnarray}
where $M_D = y_D P_{23} v_u$.


\section{Flavour textures at stabilisers} \label{sec:4}

We consider the case that $\tau_l$ and $\tau_\nu$ are fixed at stabilisers, where some analytical results can be obtained. 

Given any modular transformation $\gamma \in \Gamma_3$, a stabiliser $\tau_\gamma$ refers to some special value in the moduli space which satisfies $\gamma \tau_\gamma =\tau_\gamma$. Note that $\gamma \tau_\gamma =\tau_\gamma$ may have more than one solutions. In such case, we denote each stabiliser as $\tau_{\gamma i}$.  Stabilisers satisfy the following properties: 
\begin{itemize}
    \item A stabiliser of $\gamma$ is also a stabiliser of $\gamma^2$, i.e., $\tau_\gamma = \tau_{\gamma^2}$. 
    \item If  $\tau_\gamma$ is a stabiliser of $\gamma$,  $\gamma_1\tau_\gamma$ is a stabiliser of $\gamma_1\gamma \gamma_1^{-1}$, i.e., $\tau_{\gamma_1 \gamma \gamma_1^{-1}} = \gamma_1 \tau_\gamma$, for any given $\gamma_1 \in A_4$. 
    \item The stabiliser $\tau_\gamma$ preserves the residual symmetry generated by $\gamma$. 
\end{itemize}
There are 14 independent stabilisers in the fundamental domain of $\Gamma_3$ \cite{Ding:2019gof, deMedeirosVarzielas:2020kji, deMedeirosVarzielas:2021pug}. They and the corresponding residual symmetries are listed in Table~\ref{tab:stabilisers}, where
the former eight preserve $Z_3$ symmetries and the rest preserve $Z_2$ symmetries.

\begin{table}
	\begin{center}
	\begin{tabular}{|ll|l|}
	\hline 
	\multicolumn{2}{|l|}{Stabilisers $\tau_\gamma$} & Residual modular symmetry at $\tau_\gamma$   \\\hline\hline  
	$\tau_{T 1} = i \infty$, & $\tau_{T 2} = -\frac{1}{\omega -1}+1$ & $Z_3^T = \{1, T, T^2\}$\\
	$\tau_{STS 1} = 0$, & $\tau_{STS 2} = \omega-1$ & $Z_3^{STS} = \{1, STS, ST^2S\}$ \\
    $\tau_{ST 1} = 1$, & $\tau_{ST 2} = \omega$ & $Z_3^{ST} = \{1, ST, T^2S\}$ \\
   $\tau_{TS 1} = -1$, & $\tau_{TS 2} = \omega+1$ & $Z_3^{TS} = \{1, TS, ST^2\}$ \\\hline
   $\tau_{S 1} = i$, & $\tau_{S 2} = \frac{3}{2}+\frac{i}{2}$ & $Z_2^S=\{1, S\}$ \\
   $\tau_{T S T^2 1} =1+ i$, & $\tau_{T S T^2 2} =- \frac{1}{2}+\frac{i}{2}$ & $Z_2^{TST^2}=\{1, TST^2\}$ \\
   $\tau_{T^2 S T 1} = -1+i$, & $\tau_{T^2 S T 2} = \frac{1}{2}+\frac{i}{2}$ & $Z_2^{T^2ST}=\{1, T^2ST\}$ \\\hline\hline 
		\end{tabular}
	\end{center}
 \caption{Stabilisers and residual symmetries at stabilisers, where $\omega = (-1+\sqrt{3} i)/2 = e^{2\pi i/3} $. }\label{tab:stabilisers}
\end{table}

Stabilisers restrict flavour structures: Once $\tau$ takes a VEV at a stabiliser $\tau_\gamma$, $\Gamma_3$ is spontaneously broken, but the subgroup generated by $\gamma$ is still conserved in the relevant superpotential terms and the consequence mass terms in the Lagrangian, leading to restrictions on the mass matrix. In detail, given left- and right-handed fermions $f_L$ and $f_R$ transformed following the ordinary modular transformation 
\begin{eqnarray}
    f_L &\to& (c\tau+d)^{2k_{f_L}}\rho_{f_L}(\gamma) \, f_L \,, \nonumber\\
    f_R &\to& (c\tau+d)^{2k_{f_R}}\rho_{f_R}(\gamma) \, f_R \,,
\end{eqnarray} 
where $2k_{f_L}$ and $2k_{f_R}$ are modular weights of $f_L$ and $f_R$, and $\rho_{f_L}(\gamma)$ $\rho_{f_R}(\gamma)$ are unitary representation matrices of $\gamma$ along the representations of $f_L$ and $f_R$, respectively. Once $\tau$ takes the VEV at  $\tau_\gamma$, the mass terms, $\overline{f_L} M_f f_R + {\rm h.c.}$, is invariant under the transformation, and thus $M_f$ should satisfy the following identity
\begin{eqnarray}
   (c_\gamma\tau_\gamma+d_\gamma)^{2k_{f_R}-2k_{f_L}} \, \rho_{f_L}(\gamma)^\dag \, M_f \, \rho_{f_R}(\gamma) = M_f \,,
\end{eqnarray}
where $c_\gamma$ and $d_\gamma$ are values of $c$ and $d$ at $\gamma$.
Below, we will give a comprehensive discussion on the charged lepton and neutrino mass structure at stabilisers. 

We first discuss the viable stabilisers in the charged lepton sector. 
For modular weight at +2, The determinant of charged lepton mass matrix Eq.~\eqref{eq:M_E_2} is given by
\begin{eqnarray}
\text{Det}(M_l^*) &=&
y_e y_\mu y_\tau (Y_1+Y_2+Y_3)(Y_1^2+Y_2^2+Y_3^2 - Y_1 Y_2-Y_2 Y_3-Y_1 Y_3)
\,.
\end{eqnarray}
It is invariant under the $T$ transformation, $\tau_l \to \tau_l+1$.
Following this properties, we discuss variable stabilisers as follows.
At $\tau_l = \tau_{S1}, \tau_{S2}$, $Y$ is calculated to be $(-\sqrt{3}-2,\sqrt{3}+1,1)^{\rm T} Y_3(\tau_{S1}), (1,\sqrt{3}+1,-\sqrt{3}-2)^{\rm T} Y_1(\tau_{S2})$, respectively, where $Y_3(\tau_{S1}) \approx -0.273987$ and $Y_1(\tau_{S2}) \approx 0.54798$. The determinant $\text{Det}(M_l^*)$ vanishes in these cases, which conflicts with masses of charged leptons. 
Under the $T$ transformation, we checked $\text{Det}(M_l^*)=0$ at $\tau_l = \tau_{TST^21},\tau_{TST^22}, \tau_{T^2ST1},\tau_{T^2ST2}$. 
We are then left with the rest eight stabilisers. The Yukawa form $Y$ and charged lepton mass matrix $M_l$ at these stabilisers are calculated directly. At $\tau_l = \tau_{T1}$, $Y = (1,0,0)^{\rm T}$. We obtain a diagonal charged lepton mass matrix at $\tau_{T1}$, $M_l(\tau_{T1}) = M_l^T$, where $M_l^T$ is a notation of the following matrix invariant under the $T$ transformation,
\begin{eqnarray}
    M_l^T = \begin{pmatrix}
        y_e & 0 & 0 \\ 0 & y_\mu & 0 \\ 0 & 0 & y_\tau
    \end{pmatrix} v_d \,.
\end{eqnarray}
In other word, a modular $Z_3$ residual symmetry generated by $T$ is left unbroken in the mass matrix, which is denoted as $M_l^T$. 
At $\tau_l = \tau_{T2}$, $Y = (0,0,1)^{\rm T} Y_3(\tau_{T2})$ with $Y_3(\tau_{T2}) \approx -4.26903$. The mass matrix $M_l$ is not diagonal, $M_l(\tau_{T2}) = P^2 M_l^T$ with
\begin{eqnarray}
P = \begin{pmatrix}
        0 & 1 & 0 \\ 0 & 0 & 1 \\ 1 & 0 & 0
    \end{pmatrix} \,.
\end{eqnarray} 
The Yukawa form and charged lepton mass matrix at other stabilisers are also obtained directly. 
We classify all viable stabilisers in the charged lepton sector into two sets, labelled as case A and case B below. The modular form $Y$ and mass matrix $M_l$ at these stabilisers are given below,
\begin{eqnarray}
\label{eq:case_A}
\text{Case A}:&&\left\{\begin{array}{lll}
    \tau_l = \tau_{T1},& Y = (1,0,0)^{\rm T},& M_l = M_l^T, \\
	\tau_l = \tau_{STS1},& Y \propto (-\frac{1}{3},\frac{2}{3},\frac{2}{3})^{\rm T},& M_l = S M_l^T, \\
   	\tau_l = \tau_{ST1},& Y \propto (-\frac{1}{3},\frac{2}{3} \omega,\frac{2}{3} \omega^2)^{\rm T},&         M_l = T S T^2 M_l^T, \\
    \tau_l = \tau_{TS1},& Y \propto (-\frac{1}{3},\frac{2}{3} \omega^2,\frac{2}{3} \omega)^{\rm T},&  M_l = T^2 S T M_l^T;
\end{array}\right. \\
\label{eq:case_B}
\text{Case B}: &&\left\{\begin{array}{lll}
    \tau_l = \tau_{T2},& Y \propto (0,0,1)^{\rm T},& M_l = P^2 M_l^T, \\
	\tau_l = \tau_{STS2},& Y \propto (\frac{2}{3},\frac{2}{3},-\frac{1}{3})^{\rm T},& M_l = S P^2 M_l^T, \\
   	\tau_l = \tau_{ST2},& Y \propto (\frac{2}{3},\frac{2}{3} \omega,-\frac{1}{3} \omega^2)^{\rm T},& M_l = T S T^2 P^2 M_l^T, \\
    \tau_l = \tau_{TS2},& Y \propto (\frac{2}{3},\frac{2}{3} \omega^2,-\frac{1}{3} \omega)^{\rm T},& M_l = T^2 S T P^2 M_l^T.
\end{array}\right.
\end{eqnarray}
Here and below, without leading to any misunderstanding, we do not distinguish $S, T, \cdots$ with their triplet representation matrices in the complex basis, which are given in Appendix~\ref{sec:multiplication-rule}. 
In some of the formulas of $M_l$ above, we have dismissed an overall unphysical phase factor. 
We then discuss the modular weight at $2k_l=4$. The charged lepton mass matrix is given in Eq.~\eqref{eq:M_E_4}. Its determinant vanishes at $\tau_l = \tau_{S1}, \tau_{S2}, \tau_{TST^21},\tau_{TST^22}, \tau_{T^2ST1},\tau_{T^2ST2}$, and thus we will not consider them.
At the other stabilisers, we find that $M_l$ can also be written in similar forms as above. Again, we classify the stabilisers into two sets, and the modular form $Z$ and mass matrix $M_l$ are listed below. 
\begin{eqnarray} \label{eq:case_Ap}
\text{Case A}:&&\left\{\begin{array}{lll}
    \tau_l = \tau_{T1},& Z = (1,0,0)^{\rm T},& M_l =  M_l^T, \\
	\tau_l = \tau_{STS1},& Z \propto (-\frac{1}{3},\frac{2}{3},\frac{2}{3})^{\rm T},& M_l = S  M_l^T, \\
   	\tau_l = \tau_{ST1},& Z \propto (-\frac{1}{3},\frac{2}{3} \omega,\frac{2}{3} \omega^2)^{\rm T},& M_l = T S T^2  M_l^T, \\
    \tau_l = \tau_{TS1},& Z \propto (-\frac{1}{3},\frac{2}{3} \omega^2,\frac{2}{3} \omega)^{\rm T},& M_l = T^2 S T  M_l^T;
\end{array}\right. \\
\label{eq:case_C}
\text{Case C}:&&\left\{\begin{array}{lll}
    \tau_l = \tau_{T2},& Z \propto (0,1,0)^{\rm T},& M_l = P  M_l^T, \\
	\tau_l = \tau_{STS2},& Z \propto (\frac{2}{3},-\frac{1}{3},\frac{2}{3})^{\rm T},& M_l = S P  M_l^T,  \\
   	\tau_l = \tau_{ST2},& Z \propto (\frac{2}{3},-\frac{1}{3} \omega,\frac{2}{3} \omega^2)^{\rm T},& M_l = T S T^2 P  M_l^T, \\
    \tau_l = \tau_{TS2},& Z \propto (\frac{2}{3},-\frac{1}{3} \omega^2,\frac{2}{3} \omega)^{\rm T},& M_l = T^2 S T P M_l^T.
\end{array}\right. 
\end{eqnarray}
The first set gives the same direction for $Z$ as those for $Y$ in Eq.~\eqref{eq:case_A}. Thus the same prediction is expected as that in Eq.~\eqref{eq:case_A} and we denote them also as case A. Case C give  directions of modular forms different from cases A and B.
In summary, we obtain 12 viable $M_l$'s at different stabilisers which can be expressed in the form
\begin{eqnarray} \label{eq:M_l_general}
    M_l = T^i S^j T^{-i} P^k M_l^T
\end{eqnarray}
for $i=0,1,2$, $j=0,1$ and $k=0,1,2$. 
We have also checked that by taking $2k_l =6$, which is weight used in \cite{deMedeirosVarzielas:2021pug}, only those with $k=0$ is obtained. Even greater modular weights does not give more possibilities to the flavour structure. In all these cases, we can write $M_l$ in the unified form 
\begin{eqnarray} \label{eq:M_l}
M_l = U_l M_l^T
\end{eqnarray}
with $U_l$ a unitary matrix which can be directly read from Eq.~\eqref{eq:M_l_general}. Each of these matrices preserves a modular $Z_3$ residual symmetry of $A_4$. 

Now we consider mass matrices for neutrinos. Reminder the Majorana mass matrix for right-handed neutrinos in Eq.~\eqref{eq:M_RHN}, we discuss its behaviour at stabilisers. 
Among fourteen stabilisers, eight of them ($\tau_{T1}, \tau_{ST1}, \tau_{TS1}, \tau_{STS1}, \tau_{T2}, \tau_{ST2}, \tau_{TS2}, \tau_{STS2}$) lead to degenerate eigenvalues for neutrino masses, no matter what the values of $\mu_1$, $\mu_{1'}$ and $\mu_3$ are. Thus, we will focus on the rest six. 

At $\tau_\nu=\tau_{S1}$, we have $Y = (-\sqrt{3}-2,\sqrt{3}+1,1)^{\rm T} Y_3(\tau_{S1}) $ and
\begin{eqnarray}
X & = & 3 \left(2 \sqrt{3}+3\right) Y_3^2(\tau_{S1}) \,, \nonumber\\
X' & = & -3 \left(2 \sqrt{3}+3\right) Y_3^2(\tau_{S1}) \,, \nonumber\\
Z & = & 3 \left(2+\sqrt{3}\right) Y_3^2(\tau_{S1}) \left( 1 , 1 , 1 \right)^{\rm T} \,, 
\end{eqnarray}
where $Y_3(\tau_{S1}) \approx -0.273987$. 
It is convenient to rewrite the right-handed neutrino mass matrix in an $S$-invariant simplified form (namely satisfying $S M_R S^{\rm T} = M_R^S$)
\begin{eqnarray} \label{eq:M_nuc_S}
M_R^S = M_0 \left[
\begin{pmatrix}
2 & -1 & -1 \\ -1 & 2 & -1 \\ -1 & -1 & 2
\end{pmatrix}
+
g\begin{pmatrix}
1 & 0 & 0 \\ 
0 & 0 & 1 \\ 
0 & 1 & 0
\end{pmatrix}
+ 
g'\begin{pmatrix}
0 & 0 & 1 \\ 
0 & 1 & 0 \\ 
1 & 0 & 0
\end{pmatrix}
\right]
\,, 
\end{eqnarray}
where $M_0$ is a real and undetermined overall mass scale, and $g$ and $g'$ are two complex dimensionless parameters.
Applying the seesaw formula, we obtain the light neutrino mass matrix as
\begin{eqnarray} \label{eq:M_nu_S}
M_\nu^S = m_0 
\left[
\left(
\begin{array}{ccc}
 2 & -1 & -1 \\
 -1 & 2 & -1 \\
 -1 & -1 & 2 \\
\end{array}
\right)
+
\frac{3-g^2}{g+g'} \left(
\begin{array}{ccc}
 1 & 0 & 0 \\
 0 & 0 & 1 \\
 0 & 1 & 0 \\
\end{array}
\right)+\frac{3-g^{\prime 2}}{g+g'} \left(
\begin{array}{ccc}
 0 & 1 & 0 \\
 1 & 0 & 0 \\
 0 & 0 & 1 \\
\end{array}
\right) +
\frac{g g'+3}{g+g'} \left(
\begin{array}{ccc}
 0 & 0 & 1 \\
 0 & 1 & 0 \\
 1 & 0 & 0 \\
\end{array}
\right)
\right] \,,
\end{eqnarray}
where 
\begin{eqnarray}
    m_0 = \frac{y_D^2 v_u^2}{M_0 (9-g^2 - g^{\prime 2} +  g g')}
\end{eqnarray}
is an overall mass parameter for light neutrinos. Ref.~\cite{Novichkov:2018yse} construct the light neutrino mass via the Weinberg operator, where the third term on the right-hand side of the above equation cannot be obtained. It is convenient to check that $M_\nu^S$ preserves a modular $Z_2$ symmetry generated by $S$. 

At $\tau_\nu = \tau_{S2}$, we have $Y = (1,\sqrt{3}+1,-\sqrt{3}-2)^{\rm T} Y_1(\tau_{S2})$ and 
\begin{eqnarray}
X & = & -3 \left(2 \sqrt{3}+3\right) Y_1^2(\tau_{S2}) \,, \nonumber\\
X' & = & 3 \left(2 \sqrt{3}+3\right) Y_1^2(\tau_{S2}) \,, \nonumber\\
Z & = & 3 \left(2+\sqrt{3}\right) Y_1^2(\tau_{S2}) \left( 1 , 1 , 1 \right)^{\rm T} \,, 
\end{eqnarray}
where $Y_1(\tau_{S2}) \approx 0.54798$.
We can find that the mass matrix $M_{\nu^c}$ at $\tau_\nu = \tau_{S2}$ is exactly the same as that at $\tau_\nu = \tau_{S1}$ once we consider  parameter replacements $\mu_1 Y_3^2(\tau_{S1}) \to - \mu_1 Y_1^2(\tau_{S2})$, $\mu_{1'} Y_3^2(\tau_{S1}) \to - \mu_{1'} Y_1^2(\tau_{S2})$ and $\mu_3 Y_3^2(\tau_{S1}) \to \mu_3 Y_1^2(\tau_{S2})$. Therefore, the mass matrices in Eqs.~\eqref{eq:M_nuc_S} and \eqref{eq:M_nu_S} still apply.
Neutrino mass matrix at stabilisers $\tau_{TST^21}$, $\tau_{T^2ST1}$, $\tau_{TST^22}$ and $\tau_{T^2ST2}$ are obtained following the transformation $T^n$ (for $n=1,2$), i.e., $\tau_\nu \to T^n\tau_\nu = \tau_\nu+n$,
 $\tau_{Si} \to T^n \tau_{Si} = \tau_{T^n S T^{-n} i}$, 
\begin{eqnarray}
    M_R(\tau_{T^n S T^{-n}i}) = T^{n} M_R^S T^{n} \,.
\end{eqnarray}
Considering the seesaw formula, the light neutrino mass matrix is transformed as
\begin{eqnarray}
    M_\nu(\tau_{T^n S T^{-n}i}) = -M_D M_R^{-1}(\tau_{T^n S T^{-n}i}) M_D^{\rm T}
     = T^{n} M_\nu^S T^{n} \,,
\end{eqnarray} 
where $T M_D$ = $M_D T^2$ has been used. 
In conclusion, we have six stabilisers to give non-degenerate neutrino masses. stabilisers and mass matrices for heavy neutrinos and light neutrinos are respectively given by 
\begin{eqnarray}
\left\{\begin{array}{lll}
     \tau_\nu = \tau_{S1}, \tau_{S2}, & M_R = M_R^S, & M_{\nu} = M_{\nu}^S, \\
     \tau_\nu = \tau_{T S T^2 1}, \tau_{T S T^2 2}, & M_R = T M_R^S T, &M_{\nu} = T M_{\nu}^S T, \\
     \tau_\nu = \tau_{T^2 S T 1}, \tau_{T^2 S T 2}, & M_R = T^2 M_R^S T^2, &M_{\nu} = T^2 M_{\nu}^S T^2.
\end{array}\right.
\end{eqnarray}
The neutrino mass matrix can be analytically diagonalised via $U_\nu^\dag M_\nu U_\nu^* = {\rm diag}\{m_1, m_2, m_3\}$. Here three mass eigenvalues are solved to be
\begin{eqnarray}
m_2^2 &=& \frac{(y_D v_u)^4}{M_0^2 |g+g'|^2} \,, \nonumber\\
m_\pm^2 &=& \frac{(y_D v_u)^4}{M_0^2} \frac{1}{9+|g|^2-\text{Re}(g' g^*)+|g'|^2 \mp \sqrt{3} A} \,.
\end{eqnarray}
We can let $m_+ = m_3$ and $m_- = m_1$ for normal ordering (NO) for light neutrino masses, and $m_- = m_3$ and $m_+ = m_1$ for inverted ordering (IO) for light neutrino masses. 
The unitary matrix in NO and IO is given by
\begin{eqnarray} \label{eq:U_nu}
U_\nu &=& T^n \left(
\begin{array}{ccc}
 \frac{\sqrt{\frac{4}{6}}C_-}{B_-} & \sqrt{\frac{1}{3}} & \frac{\sqrt{\frac{4}{6}}C_+}{B_+} \\
 \frac{-\sqrt{\frac{1}{6}}C_- -\sqrt{\frac{1}{2}}D}{B_-} & \sqrt{\frac{1}{3}} & \frac{-\sqrt{\frac{1}{6}}C_+ -\sqrt{\frac{1}{2}}D}{B_+} \\
 \frac{-\sqrt{\frac{1}{6}}C_- +\sqrt{\frac{1}{2}}D}{B_-} & \sqrt{\frac{1}{3}} & \frac{-\sqrt{\frac{1}{6}}C_+ +\sqrt{\frac{1}{2}}D}{B_+} \\
\end{array}
\right) P_+ \quad\text{for NO}, \nonumber\\
U_\nu &=& T^n \left(
\begin{array}{ccc}
 \frac{\sqrt{\frac{4}{6}}C_+}{B_+} & \sqrt{\frac{1}{3}} & \frac{\sqrt{\frac{4}{6}}C_-}{B_-} \\
 \frac{-\sqrt{\frac{1}{6}}C_+ -\sqrt{\frac{1}{2}}D}{B_+} & \sqrt{\frac{1}{3}} & \frac{-\sqrt{\frac{1}{6}}C_- -\sqrt{\frac{1}{2}}D}{B_-} \\
 \frac{-\sqrt{\frac{1}{6}}C_+ +\sqrt{\frac{1}{2}}D}{B_+} & \sqrt{\frac{1}{3}} & \frac{-\sqrt{\frac{1}{6}}C_- +\sqrt{\frac{1}{2}}D}{B_-} \\
\end{array}
\right) P_- \quad\text{for IO}, \label{eq:U_nu}
\end{eqnarray}
respectively for $n=0,1,2$, where $P_\pm = {\rm diag}\{e^{i  {\rm Arg} (E_\mp)/2}, e^{i {\rm Arg} (g+g')/2 }, e^{i {\rm Arg} (E_\pm)/2}\}$ and
\begin{eqnarray} \label{eq:ABCDE}
A &=& \sqrt{6(\text{Re} g)^2 + 6(\text{Re} g')^2 + 6(\text{Re} g - \text{Re} g')^2 + [\text{Im} (g' g^*)]^2} \,, \nonumber\\
B_\pm &=& \sqrt{C_\pm^2 + |D|^2} \,, \nonumber\\
C_\pm &=& \sqrt{3} \text{Re} (g' -2g) \pm A \,, \nonumber\\
D &=& 3 \text{Re} g' +i \text{Im} (g' g^*) \,, \nonumber\\
E_\pm &=& 3+g-\frac{g'}{2} - \frac{\sqrt{3} g' D^*}{2 C_\pm} \,.
\end{eqnarray}

\begin{figure}[t!]
\begin{center}
    \includegraphics[width=.6\textwidth]{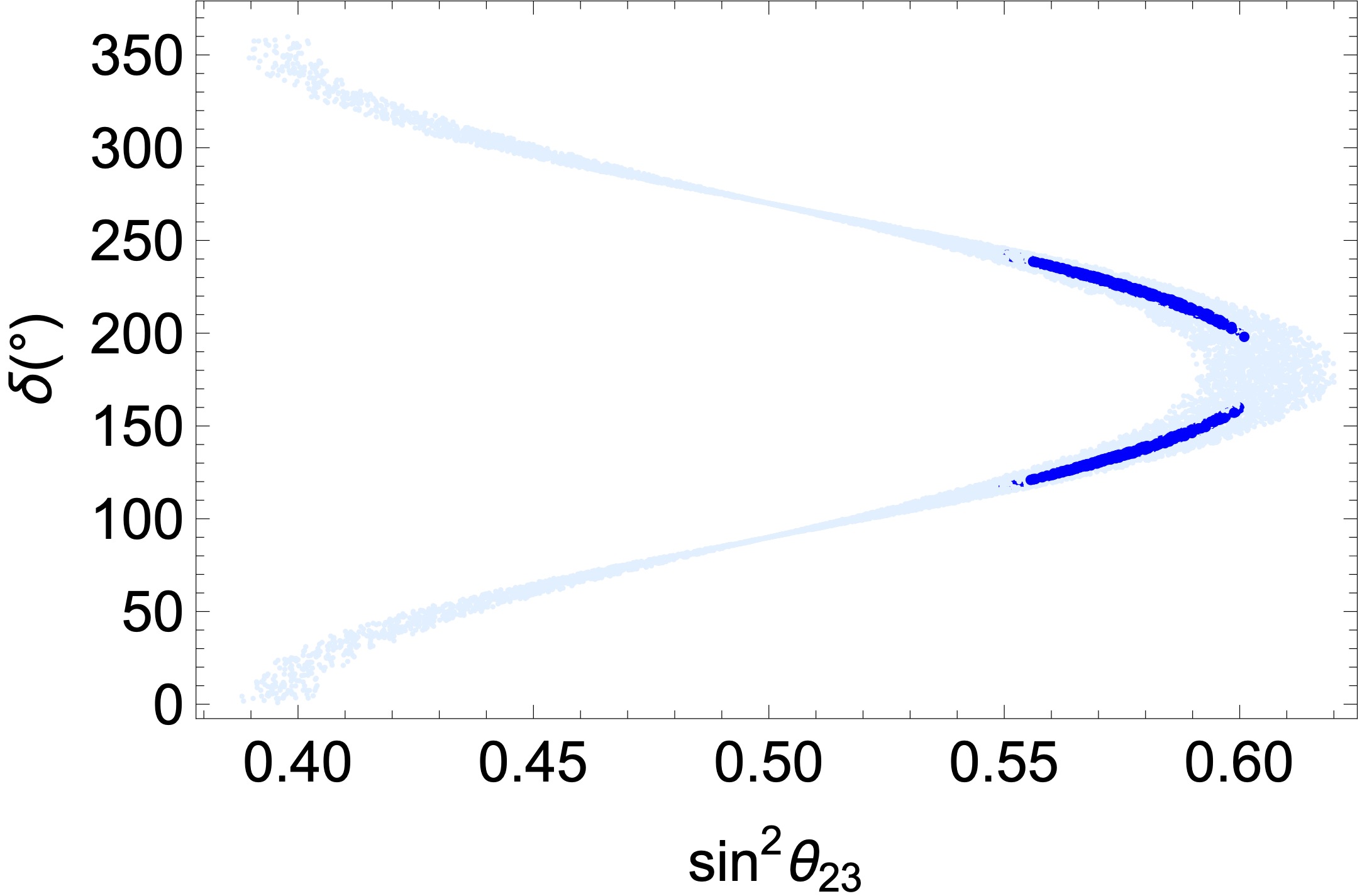}
\end{center}
	\caption{The prediction of $\delta$ versus $\sin^2\theta_{23}$ for neutrino mass in the NO. Points for $\chi^2 <10$ (blue) and $\chi^2 < 100$ (light blue) are listed. The IO gives the same distribution and thus is not presented separately.} \label{eq:theta23_delta}
\end{figure}

We discuss the property of the lepton flavour mixing matrix, which is a combination of unitary matrices $U_l$ and $U_\nu$, $U = U_l^\dag U_\nu$. The mixing matrix is parameterised in the form 
\begin{eqnarray}
    U = P_l \left(
\begin{array}{ccc}
 c_{12} c_{13} & c_{13} s_{12} & e^{-i \delta } s_{13} \\
 -c_{23} s_{12}-c_{12} e^{i \delta } s_{13} s_{23} & c_{12} c_{23}-e^{i \delta } s_{12} s_{13} s_{23} & c_{13} s_{23} \\
 s_{12} s_{23}-c_{12} c_{23} e^{i \delta } s_{13} & -c_{12} s_{23}-c_{23} e^{i \delta } s_{12} s_{13} & c_{13} c_{23} \\
\end{array}
\right) P_\nu \,,
\end{eqnarray}
where $c_{ij} = \cos \theta_{ij}$, $s_{ij} = \sin \theta_{ij}$ with three mixing angles $\theta_{12}$, $\theta_{13}$ and $\theta_{23}$, $P_\nu = {\rm diag}\{1, e^{i\alpha_{21}/2}, e^{i\alpha_{31}/2} \}$ parametrises the contribution of two Majorana phases $\alpha_{21}$ and $\alpha_{31}$, and $P_l = {\rm diag} \{ e^{i\beta_1}, e^{i\beta_2}, e^{i\beta_3} \}$ refers to the three unphysical phases which can be rotated away by a phase redefinition of right-handed charged leptons. Given general forms of $U_l$ and $U_\nu$ from Eqs.~\eqref{eq:M_l} and \eqref{eq:U_nu}, one can prove that 
\begin{eqnarray}
|U| = \begin{pmatrix} \times & \frac{1}{\sqrt{3}} & \times \\ \times & \frac{1}{\sqrt{3}} & \times \\ \times & \frac{1}{\sqrt{3}} & \times \\
\end{pmatrix} \,.
\end{eqnarray}
Namely, the TM2 mixing, as well as the corresponding sum rule
\begin{eqnarray}
    \sin^2 \theta_{12} = \frac{1}{3(1-\sin^2 \theta_{13})}
\end{eqnarray}
is always predicted at any available stabilisers in the framework of two modular $A_4$ symmetries.

We show in Fig.~\ref{eq:theta23_delta} the numerical correlation between $\theta_{23}$ and the Dirac CP phase $\delta$. Here we used a simple $\chi^2$ analysis with the $\chi^2$ function defined as
\begin{eqnarray}
\chi^2 = \sum_{i=1}^{4} \left(\frac{P_i - {\rm BF}_{i} }{\sigma_i }\right)^2 \, .
\end{eqnarray}
Here $P_i$ (for $i=1,2,3,4$) $\in \{\theta_{12},\theta_{23},\theta_{13}, \alpha\}$ with $\alpha = \Delta m_{21}^2/\Delta m_{31}^2$ for NO, and $\Delta m_{21}^2/ \Delta m_{32}^2$ for IO. ${\rm BF}_{i}$ and $\sigma_i$ are the best fit (BF) value and the $1\sigma$ error of $P_i$, which are obtained from NuFIT v5.2 \cite{Esteban:2020cvm,web_link}.
We let $\tau_l$ and $\tau_\nu$ be fixed at stabilisers we discussed above and scan parameters $g$ and $g'$ randomly in the complex plane with the absolute value restricted in $(0,10]$. Then, we present the prediction of the Dirac phase $\delta$ versus $\sin^2\theta_{23}$ for both NO and IO. Using the Type-I seesaw mechanism, we found that such prediction will always be same, no matter what the mass order, the values of $\tau_l$ and $\tau_\nu$ or the weight $2k_l$ are. Therefore all values are put in on single figure. The independence of mass ordering is simply seen in Eq.~\eqref{eq:U_nu} by changing signs for $g$ and $g'$ in Eq.~\eqref{eq:ABCDE}. The prove of insensitivity to cases A, B, and C are not that straightforward. Instead, we first write out the expression of $\theta_{23}$ and $\delta$ in terms of $C_+$ and $D$ in the normal ordering for case A as
\begin{eqnarray} \label{eq:theta23_delta_case_A}
&&\sin^2\theta_{23} = \frac{1}{2}+\frac{{\rm Re}D}{|D|}\frac{\sqrt{3}C_+|D|}{C_+^2+3|D|^2} \nonumber\\
&&\sin \delta = - \frac{{\rm Im}D}{|D|}\frac{C_+^2 + 3|D|^2}{|C_+^2-3 D^2|} 
\end{eqnarray} 
These values are directly calculated by identify the mixing matrix $U$ to $U_\nu$ in Eq.~\eqref{eq:U_nu} up to diagonal phase matrix.
In Case B, the mixing matrix is different from $U$ by including $P^2$ on the right hand side. Expressions of $\sin^2\theta_{23}$ and $\sin\delta$ in terms of $C_+$ and $D$ is more complicated than those in Eq.~\eqref{eq:theta23_delta_case_A}. However, with a reparametrisation,
\begin{eqnarray} \label{eq:theta23_delta_case_A}
&&C_+' = -\frac{1}{2}\sqrt{C_+^2-3|D|^2-2\sqrt{3}C_+ {\rm Re}D} \nonumber\\
&&D' = \frac{-\frac{\sqrt{3}}{2}(C_+^2-|D|^2) + C_+ {\rm Re}D +2 i C_+ {\rm Im}D}{\sqrt{C_+^2-3|D|^2-2\sqrt{3}C_+ {\rm Re}D}} \,,
\end{eqnarray} 
we have proved that $\sin^2\theta_{23}$ and $\sin\delta$ are expressed in a similar forms as Eq.~\eqref{eq:theta23_delta_case_A} with $C_+$ and $D$ replaced by $C_+'$ and $D'$, respectively. Case C can be proved similarly, which will not be repeated. In all three cases, we observe that the $\mu$-$\tau$ permutation symmetry in the mixing matrix, i.e., $|U_{\mu i}| = |U_{\tau i}|$ for $i=1,2,3$, which predicts $\theta_{23} = 45^\circ$ and $\delta = \pm 90^\circ$ if $\theta_{23}\neq 0$\cite{Xing:2008fg,Xing:2014zka}, is included in the parameter space. Setting ${\rm Re} D = 0$ in case A (or ${\rm Re} D' \propto \frac{\sqrt{3}}{2}(C_+^2-|D|^2) + C_+ {\rm Re}D = 0$ in case B), we arrive at the $\mu$-$\tau$ permutation symmetry. Furthermore, we obtain an approximate sum rule between $\theta_{23}$ and $\delta$ as 
\begin{eqnarray}
\frac{2}{\sin^2\theta_{13}} \Big( \sin^2\theta_{23} - \frac{1}{2} \Big)^2 + \sin^2 \delta \approx 1 \,.
\end{eqnarray}
This can be proved, e.g., in case A, by using the approximation $\sin^2\theta_{13}\approx \frac{2C_+^2}{3|D|^2}$ and $\sin \delta \approx - \frac{{\rm Im}D}{|D|}$.

\begin{figure}[t!]
  \begin{subfigure}[t]{\textwidth}
  \begin{center}
  \includegraphics[height=.28\textwidth]{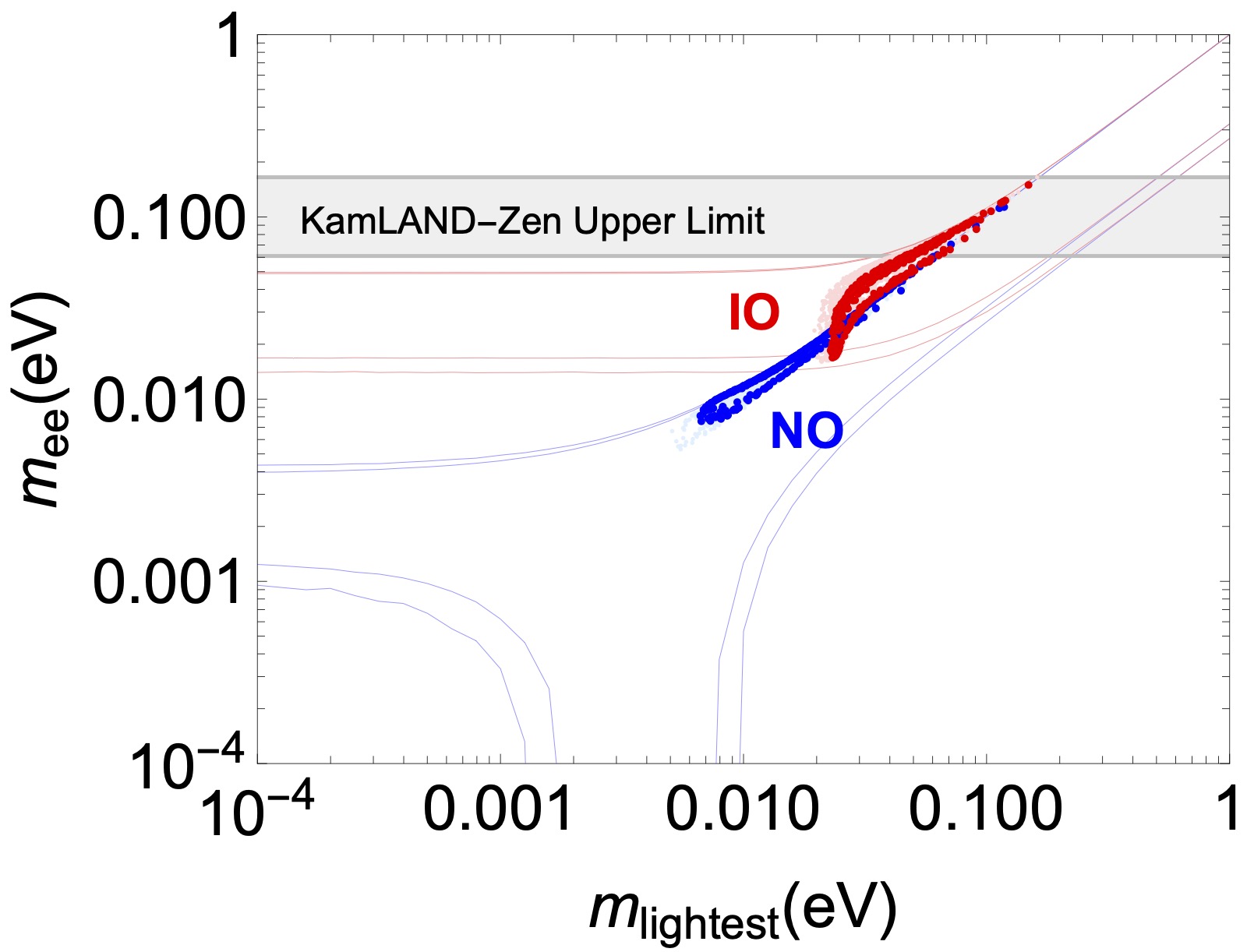}
  \includegraphics[height=.28\textwidth]{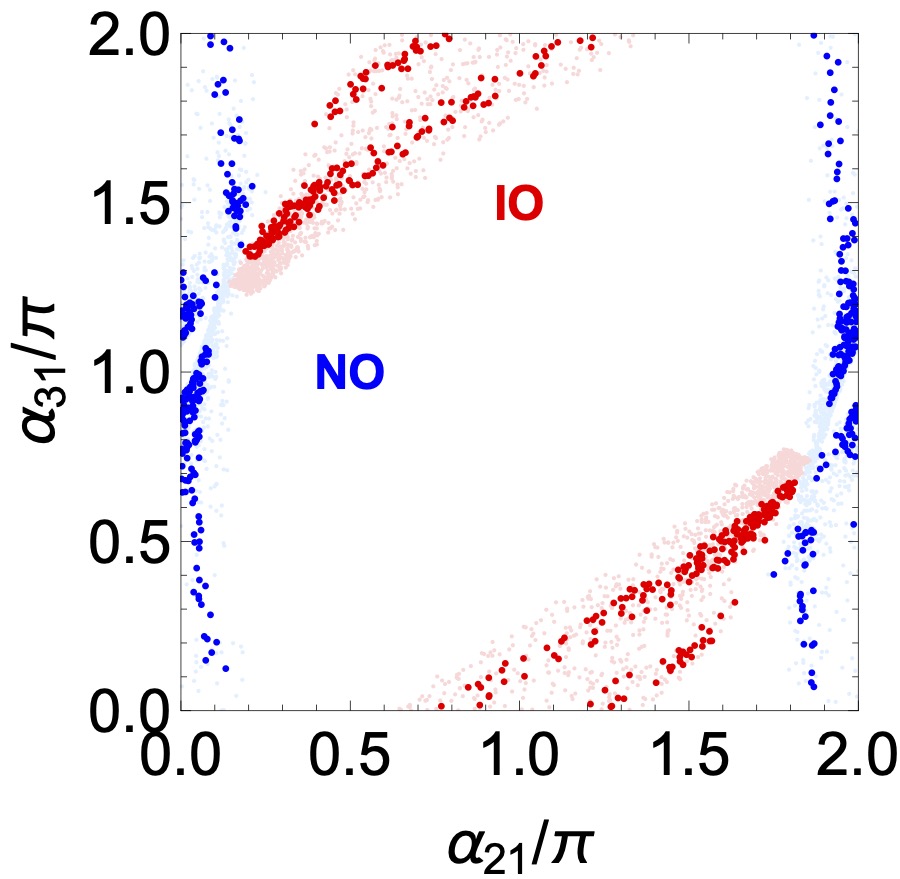}      
  \includegraphics[height=.28\textwidth]{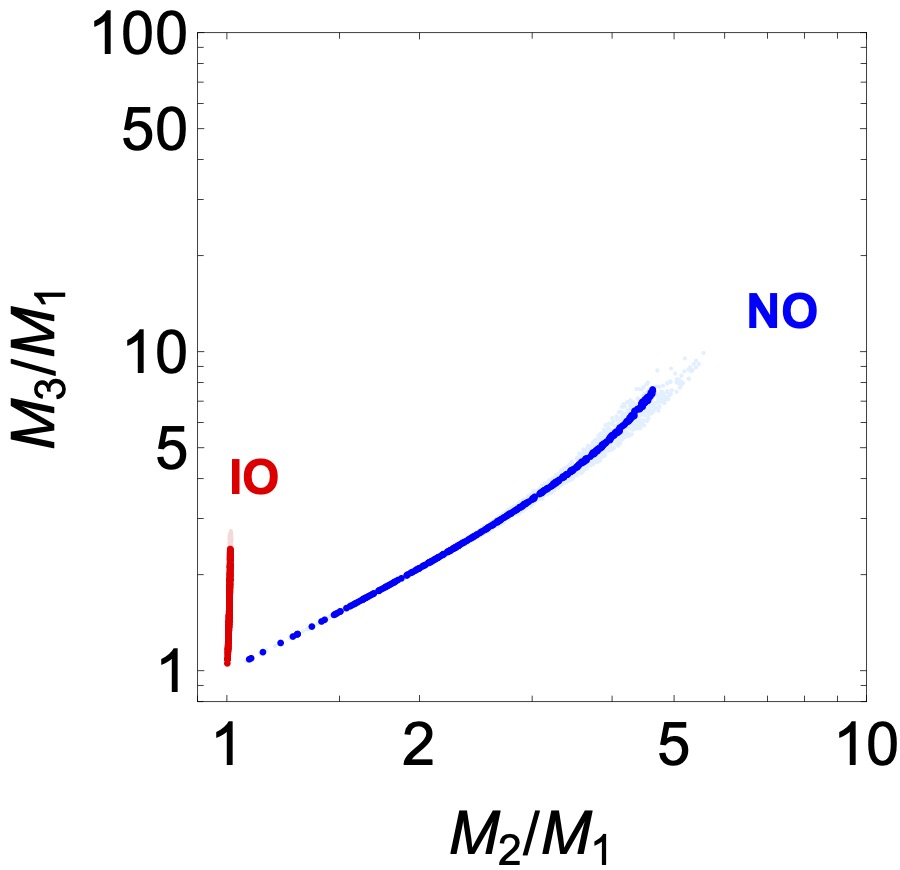}
  \caption{$\tau_l$ in case A and $\tau_\nu$ at any of the six available stabilisers.}
  \end{center} 
  \end{subfigure}
  \begin{subfigure}[t]{\textwidth}
  \begin{center}
  \includegraphics[height=.28\textwidth]{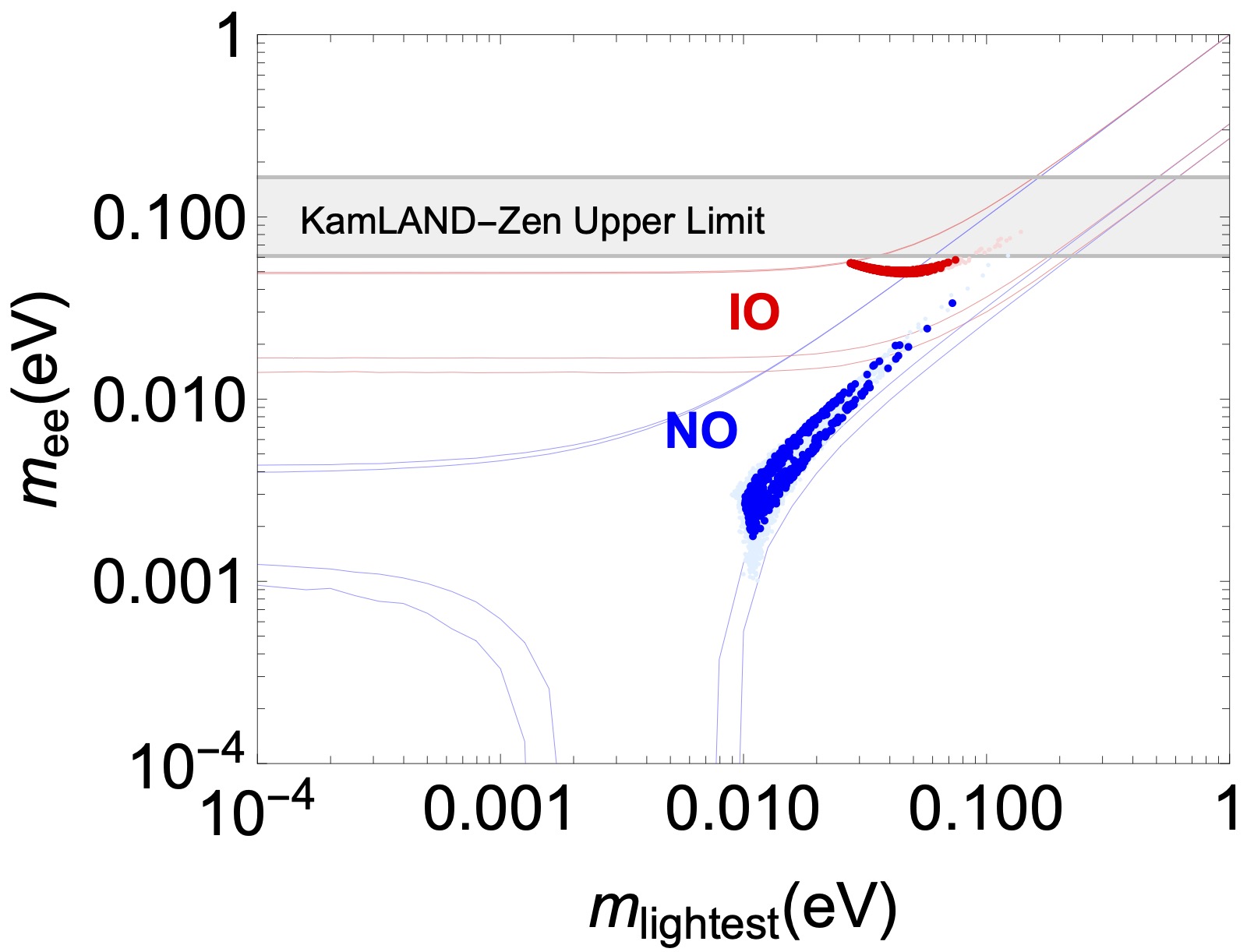}
  \includegraphics[height=.28\textwidth]{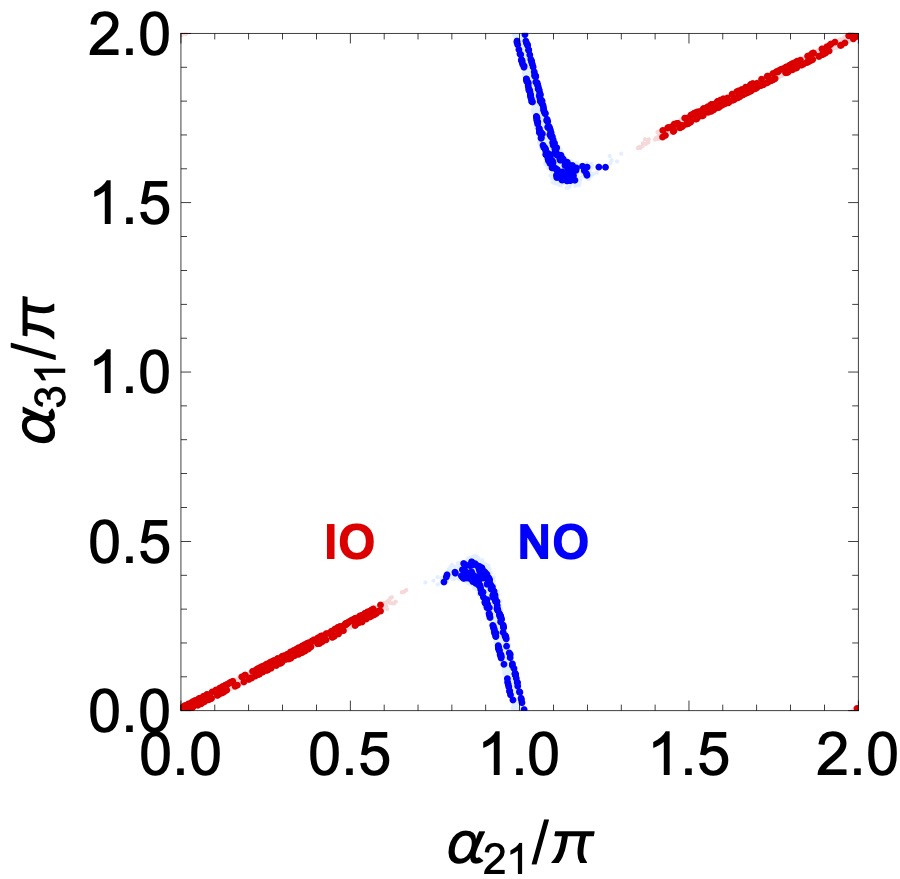}
  \includegraphics[height=.28\textwidth]{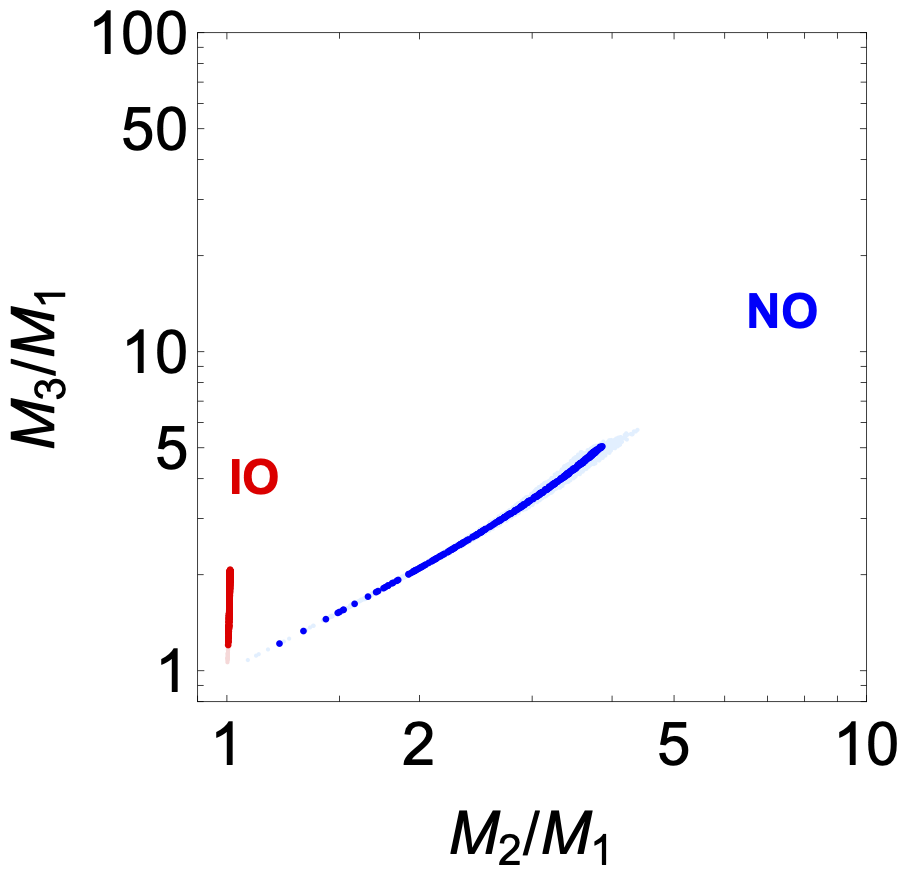}
  \caption{$\tau_l$ in case B and $\tau_\nu$ at any of the six available stabilisers.}
  \end{center} 
  \end{subfigure}
  \begin{subfigure}[t]{\textwidth}
  \begin{center}
  \includegraphics[height=.28\textwidth]{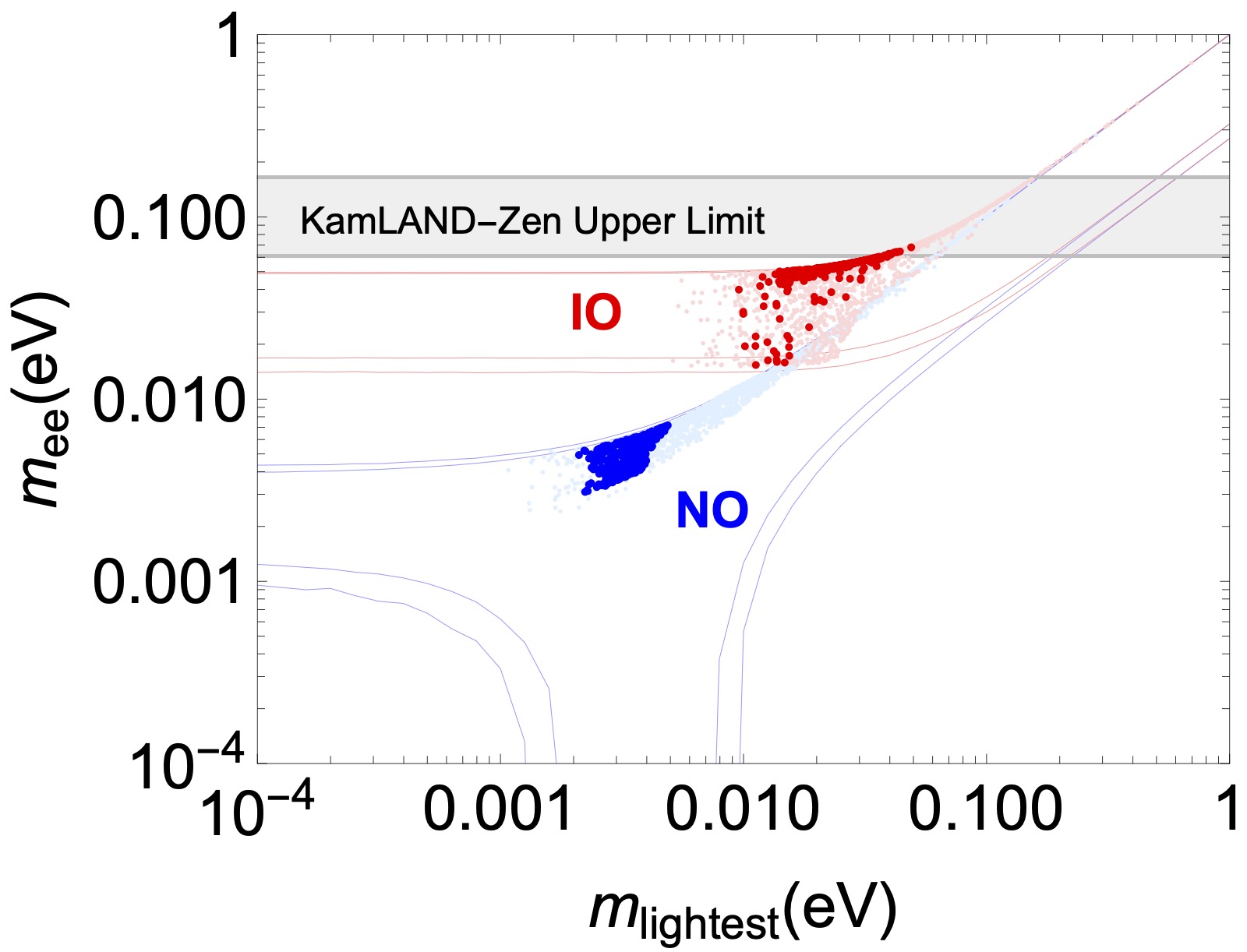}
  \includegraphics[height=.28\textwidth]{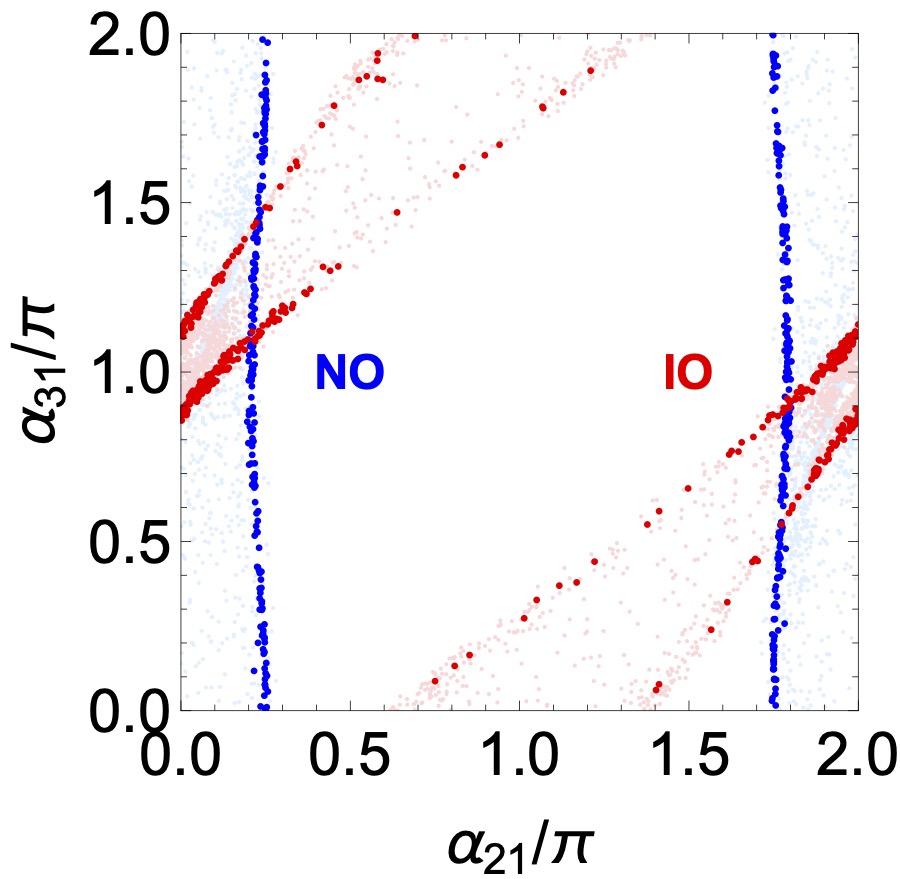}
  \includegraphics[height=.28\textwidth]{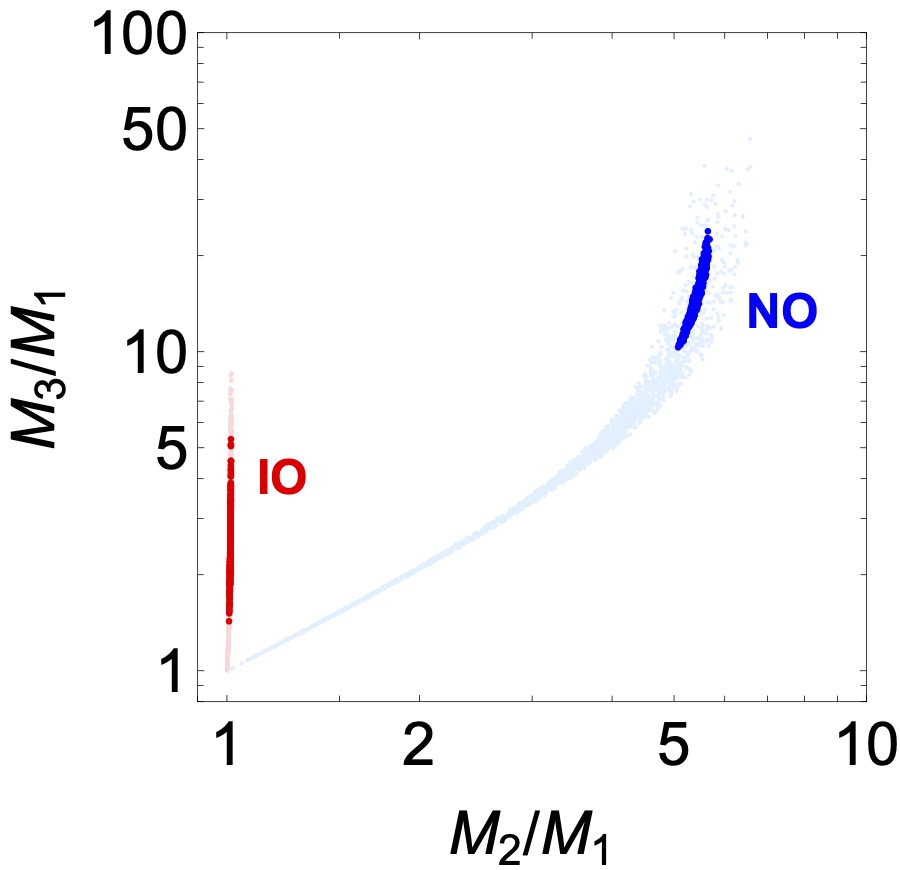}
  \caption{$\tau_l$ in case C and $\tau_\nu$ at any of the six available stabilisers.}
  \end{center} 
  \end{subfigure}
  \caption{Predictions of mass parameters $m_{ee}$ vs $m_{\rm lightest}$, Majorana phases $\alpha_{31}$ vs $\alpha_{21}$, and right-handed neutrino mass ratios $M_2/M_1$ vs $M_3/M_1$ in three cases. $\tau_l$ and $\tau_\nu$ are fixed at stabilisers mentioned in subcaptions. Points for $\chi^2 <10$ (blue for NO, red for IO) and $\chi^2 < 100$ (light blue for NO, light red for IO) are listed.}
  \label{fig:mass_parameters}
\end{figure}

We also present predictions for $m_{ee}$ which contributes to neutrinoless double beta decay vs the lightest neutrino $m_{\rm lightest}$ which is either $m_1$ for NO or $m_3$ for IO, as well as the Majorana phases $\alpha_{21}$ vs $\alpha_{31}$ in Fig.~\ref{fig:mass_parameters}. 
Different from the prediction of mixing angles and the Dirac CP phase, we obtain different predictions for these mass parameters. As discussed in Eqs.~\eqref{eq:case_A}-\eqref{eq:case_C}, three sets of lepton masses are classified into cases A, B, and C, respectively. These three cases lead to three different regions of predictions, while $\tau_\nu$ taking any of the six available stabilisers ($\tau_{S1}, \tau_{S2}, \tau_{TST^21}, \tau_{TST^22}, \tau_{T^2ST1}, \tau_{T^2ST2}$) does not lead to a difference. To see how the difference comes from, we write the light neutrino mass matrix in the flavour basis, $\widetilde{M}_\nu = U_l^\dag M_\nu U_l^*$. In cases A, B, and C, $\widetilde{M}_\nu$ is respectively expressed in the form
\begin{eqnarray}
\widetilde{M}_\nu^A &=& m_0 T^m
\left[
\pm\left(
\begin{array}{ccc}
 2 & -1 & -1 \\
 -1 & 2 & -1 \\
 -1 & -1 & 2 \\
\end{array}
\right)
+
\frac{3-g^2}{g+g'} \left(
\begin{array}{ccc}
 1 & 0 & 0 \\
 0 & 0 & 1 \\
 0 & 1 & 0 \\
\end{array}
\right)+\frac{3-g^{\prime 2}}{g+g'} \left(
\begin{array}{ccc}
 0 & 1 & 0 \\
 1 & 0 & 0 \\
 0 & 0 & 1 \\
\end{array}
\right) +
\frac{g g'+3}{g+g'} \left(
\begin{array}{ccc}
 0 & 0 & 1 \\
 0 & 1 & 0 \\
 1 & 0 & 0 \\
\end{array}
\right)
\right] T^m \,, \nonumber\\
\widetilde{M}_\nu^B &=& m_0 T^m
\left[
\pm\left(
\begin{array}{ccc}
 2 & -1 & -1 \\
 -1 & 2 & -1 \\
 -1 & -1 & 2 \\
\end{array}
\right)
+
\frac{3-g^2}{g+g'} \left(
\begin{array}{ccc}
 0 & 1 & 0 \\
 1 & 0 & 0 \\
 0 & 0 & 1 \\
\end{array}
\right)+\frac{3-g^{\prime 2}}{g+g'} \left(
\begin{array}{ccc}
 0 & 0 & 1 \\
 0 & 1 & 0 \\
 1 & 0 & 0 \\
\end{array}
\right) +
\frac{g g'+3}{g+g'} \left(
\begin{array}{ccc}
 1 & 0 & 0 \\
 0 & 0 & 1 \\
 0 & 1 & 0 \\
\end{array}
\right)
\right] T^m \,, \nonumber\\
\widetilde{M}_\nu^C &=& m_0 T^m
\left[
\pm\left(
\begin{array}{ccc}
 2 & -1 & -1 \\
 -1 & 2 & -1 \\
 -1 & -1 & 2 \\
\end{array}
\right)
+
\frac{3-g^2}{g+g'} \left(
\begin{array}{ccc}
 0 & 0 & 1 \\
 0 & 1 & 0 \\
 1 & 0 & 0 \\
\end{array}
\right)+\frac{3-g^{\prime 2}}{g+g'} \left(
\begin{array}{ccc}
 1 & 0 & 0 \\
 0 & 0 & 1 \\
 0 & 1 & 0 \\
\end{array}
\right) +
\frac{g g'+3}{g+g'} \left(
\begin{array}{ccc}
 0 & 1 & 0 \\
 1 & 0 & 0 \\
 0 & 0 & 1 \\
\end{array}
\right)
\right] T^m \,. \nonumber\\
\end{eqnarray} 
Here, $T^m$ (for $m=0,1,2$) and the $\pm$ sign on the right-hand side come from the different choices of stabilisers in corresponding sets for $\tau_l$ and $\tau_\nu$. They do not lead to any difference on the prediction because: 1) $T^m$ gives only an unphysical phase rotation on the left-hand side of the mixing matrix, and 2) The $\pm$ signs are connected via $(g, g') \leftrightarrow (-g, -g')$, and thus scanning $g$ and $g'$ in the complex plane does not distinguish the sign difference. Then, the only difference comes from the different combination between the three permutation matrices and the coefficients $g$ and $g'$ on the right-hand side of the formula.
Numerically, we have checked that the distributions of $m_{ee}$ vs $m_{\rm lightest}$ are quite different in three cases. In the NO, predicted values of $m_1$ are mainly distributed between 0.01 eV and 0.05 eV for case A and case B, and around 0.004 eV for case C, and values of $m_{ee}$ between 0.001 eV and 0.05 eV for case A, between 0.002 and 0.01 for case B, around 0.005 eV for case C. In the IO, predicted values of $m_3$ are mainly distributed between 0.01 eV and 0.1 eV for all three cases, and values of $m_{ee}$ between 0.012 eV and 0.1 eV for case A, around 0.05 eV for case B, between 0.012 eV and 0.05 eV for case C. 
All these cases show $m_{ee}$ above meV, not falling into the ``dark well'' region \cite{Xing:2015zha, Cao:2019hli, Denton:2023hkx}. 
For each cases, we show benchmark points with minimal $\chi^2$ values in Table~\ref{tab:benchmarks}.

We have also checked the prediction of right-handed neutrino mass spectrum. According to the seesaw formula and $M_D\propto P_{23}$, $M_1$, $M_2$ and $M_3$ can be obtained via a permutation of $(y_D v_u)^2/m_1$, $(y_D v_u)^2/m_2$ and $(y_D v_u)^2/m_3$. For NO of light neutrino masses, the mass ratios satisfy $M_2/M_1 = m_3/m_2$ and $M_3/M_1 = m_3/m_1$; for IO, $M_2/M_1 = m_2/m_1$ and $M_3/M_1 = m_2/m_3$. Here we used the convention $M_1 < M_2 < M_3$ for both NO and IO. 
As $\Delta m_{21}^2$, $\Delta m_{31}^2$ and $\Delta m_{32}^2$ have been measures, we have correlations for the right-handed neutrino mass ratios
\begin{eqnarray}
    \frac{M_3}{M_1} &=& \frac{M_2}{M_1} \sqrt{\frac{ 1-\alpha }{ 1 - \alpha (\frac{M_2}{M_1})^2 }} \hspace{9mm} \quad\text{for NO}\,,\nonumber\\
    \frac{M_3}{M_1} &=& \frac{M_2}{M_1} \sqrt{\frac{\alpha}{(1+\alpha)(\frac{M_2}{M_1})^2 - 1}} \quad\text{for IO}
\end{eqnarray}
in all three cases. However, the predicted regions of $M_2/M_1$ and $M_3/M_1$ are different, as seen in Fig.~\ref{fig:mass_parameters}. In particular, case C prefers a mass hierarchy larger than the other two cases, $M_2/M_1 \simeq (5,6)$ and $M_3/M_1 \simeq (10,30)$. 

\begin{table}
 	\begin{center}
		\begin{tabular}{|c|c|c|c|c|c|c|}
			\hline 
   & \multicolumn{3}{c|}{Normal ordering} & 
   \multicolumn{3}{c|}{Inverted ordering} \\\cline{2-7}
   & Case A & Case B & Case C
   & Case A & Case B & Case C \\\hline
   $\chi^2$ & $8.80$ & $8.80$ & $8.83$ & $8.75$ & $8.76$ & $8.75$ \\
   $g$  & $5.08 e^{i 6.04}$ & $1.76 e^{i 3.31}$ & $1.39 e^{i 2.37}$ & $3.91 e^{i 1.81}$ & $0.78 e^{i 0.15}$ & $1.92 e^{i 6.23}$ \\ 
   $g'$ & $1.42 e^{i 3.49}$ & $2.46 e^{i 5.97}$ & $2.66 e^{i 3.28}$ & $0.37 e^{i 5.48}$ & $1.63 e^{i 3.99}$ & $5.55 e^{i 5.74}$ \\
   $\theta_{12}$ & $35.72^\circ$ & $35.72^\circ$ & $35.72^\circ$ & $35.73^\circ$ & $35.73^\circ$ & $35.73^\circ$ \\
   $\theta_{23}$ & $49.28^\circ$ & $49.11^\circ$ & $49.35^\circ$ & $49.44^\circ$ & $49.57^\circ$ & $49.60^\circ$ \\
   $\theta_{13}$ & $8.58^\circ$ & $8.57^\circ$ & $8.58^\circ$ & $8.60^\circ$ & $8.59^\circ$ & $8.59^\circ$ \\
   $\delta$ & $134^\circ$ & $228^\circ$ & $225^\circ$ & $224^\circ$ & $222^\circ$ & $222^\circ$ \\
   $\alpha$ & $0.0295$ & $0.0295$ & $0.0296$ & $-0.0297$ & $-0.0298$ & $-0.0297$ \\
   $m_{\rm lightest}$ & $0.0126$ eV & $0.0108$ eV & $0.0030$ eV & $0.0658$ eV & $0.0329$ eV & $0.0231$ eV \\
   $m_{ee}$ & $0.0139$ eV & $0.0022$ eV & $0.0036$ eV & $0.0759$ eV & $0.0517$ eV & $0.0535$ eV \\
   $\alpha_{21}$ & $354.9^\circ$ & $183.6^\circ$ & $319.8^\circ$ & $313.6^\circ$ & $303.0^\circ$ & $1.6^\circ$ \\
   $\alpha_{31}$ & $209.8^\circ$ & $335.8^\circ$ & $248.6^\circ$ & $105.5^\circ$ & $329.5^\circ$ & $157.9^\circ$ \\
   $M_2/M_1$ & $3.386$ & $3.706$ & $5.497$ & $1.005$ & $1.011$ & $1.012$ \\
   $M_3/M_1$ & $4.101$ & $4.732$ & $16.527$ & $1.256$ & $1.818$ & $2.384$ \\
			\hline\hline  		
		\end{tabular}
	\end{center}
 \caption{Inputs and predictions for points with minimal $\chi^2$ values for each cases in our scan. In case A, we take $\tau_l = \tau_{T1}$, in cases B and C, we take $\tau_l = \tau_{T2}$, and $\tau_\nu = \tau_{S1}$ in all cases. $\alpha = \Delta m_{21}^2/\Delta m_{31}^2$ for NO, and $\Delta m_{21}^2/ \Delta m_{32}^2$ for IO. }\label{tab:benchmarks}
\end{table}

\section{Summary} \label{sec:5}

We discussed lepton flavour mixing in modular $A_4^l \times A_4^\nu$ symmetries. With the help of a bi-triplet scalar, two $A_4$ symmetries are spontaneously broken to a single $A_4$ after the scalar gains the VEV. The lower-energy theory appears as an effective flavour model in one modular $A_4$ symmetry evolving  two moduli fields, $\tau_l$ contributing to charged lepton Yukawa couplings and $\tau_\nu$ contributing to right-handed neutrino mass matrix. 

We gave a comprehensive study on lepton flavour mixing for moduli fields fixed at stabilisers. We let both moduli fields $\tau_l$ and $\tau_\nu$ scan among all 14 stabilisers of $\Gamma_3$, 8 of them preserving modular $Z_3$ symmetries and the other 6 preserving modular $Z_2$ symmetries. In order to generate correct charged lepton mass spectrum, $\tau_l$ has to take a $Z_3$-invariant stabiliser. $\tau_\nu$ should be fixed at a $Z_2$ stabiliser to avoid degenerate masses for light neutrinos. We have proved that the TM2 mixing pattern is always generated at any of these stabilisers. However the predicted region of $m_{ee}$ vs $m_{\rm lightest}$, as well as the correlation between two Majorana phases $\alpha_{21}$ and $\alpha_{31}$, are different, and classified into three cases for light neutrino masses in either normal ordering or inverted ordering.

\vspace{0.5cm}
\noindent

{\bf Acknowledgement}  

The work was partly supported by National Natural Science Foundation of China under Grant No. 12205064 and Zhejiang Provincial Natural Science Foundation of China under Grant No. LDQ24A050002. 


\appendix

\section*{Appendix}

\section{Multiplication rule of $A_4$ group}
\label{sec:multiplication-rule}

\begin{table}[t!]
\begin{center}
\begin{tabular}{|c|ccccc|}
\hline\hline
& $\mathbf{1}$ & $\mathbf{1'}$ & $\mathbf{1''}$ & $\mathbf{3}$ in complex basis &  $\mathbf{3}$ in real basis \\\hline
$T$ & 1 & $\omega$ & $\omega^2$ & 
$\left(
\begin{array}{ccc}
 1 & 0 & 0 \\
 0 & \omega & 0 \\
 0 & 0 & \omega ^2  \\
\end{array}
\right)$ & 
$\begin{pmatrix} 0 & 1 & 0 \\ 0 & 0 & 1 \\ 1 & 0 & 0 \end{pmatrix}$ \\

$S$ & 1 & 1 & 1 &
$\frac{1}{3} \left(
\begin{array}{ccc}
 -1 & 2 & 2 \\
 2 & -1 & 2 \\
 2 & 2 & -1 \\
\end{array}
\right)$ &
$\begin{pmatrix} 1 & 0 & 0 \\ 0 & -1 & 0 \\ 0 & 0 & -1 \end{pmatrix}$
\\ \hline\hline

\end{tabular}
\caption{\label{tab:rep_matrix} The representation matrices for the $A_4$ generators $T$ and $S$ used in the main text. $T$ and $S$ in the triplet complex basis $\rho_{\bf 3}$ and the triplet real basis $\tilde{\rho}_{\bf 3}$ are connected via  Eq.~\eqref{eq:basis_transformation}.}
\end{center}
\end{table}

$A_4$ is the group of even permutations of four objects. It is a subgroup of $\overline{\Gamma}$, with generators satisfying $S^2 = (ST)^3 = T^3 = 1$. 
It contains 12 elements. The latter are classified into four conjugacy classes 
\begin{eqnarray}
&&C_1 = \{ 1 \} \,,\nonumber\\
&&C_2 = \{ S, TST^2, T^2S T\} \,, \nonumber\\
&&C_3 = \{ T, STS, ST, TS\} \,, \nonumber\\
&&C_3' = \{ T^2, ST^2S, T^2S, ST^2\} \,.
\end{eqnarray}
$A_4$ contains 4 irreducible representations, ${\bf 1}$, ${\bf 1'}$, ${\bf 1''}$ and ${\bf 3}$. The representations follow the multiplicities
\begin{eqnarray}
& {\bf 1} \otimes {\bf 1} = {\bf 1} \ , \qquad
{\bf 1'} \otimes {\bf 1'} = {\bf 1''} \ , \qquad
{\bf 1''} \otimes {\bf 1''} = {\bf 1'} \ , \qquad
{\bf 1'} \otimes {\bf 1''} = {\bf 1} \  . \nonumber\\
&\mathbf{1^{(\prime,\prime\prime)}}\otimes\mathbf{3}=\mathbf{3} \ , \qquad
\mathbf{3}\otimes\mathbf{3}=\mathbf{1}\oplus \mathbf{1}' \oplus \mathbf{1}'' \oplus\mathbf{3}_S\oplus\mathbf{3}_A \ .
\end{eqnarray}

Generators in the representations are given by matrices in Table~\ref{tab:rep_matrix}. We listed the triplet representation of $T$ and $S$ in both the complex basis $\rho_{\bf 3}$ and real basis $\tilde{\rho}_{\bf 3}$. The former one is widely used to derive the flavour mixing and the latter is more convenient to derive the VEV of the bi-triplet scalar $\Phi$ as discussed in appendix \ref{sec:vacuum-alignments}. The two bases are connected via a unitary $U_\omega$, 
\begin{eqnarray} \label{eq:basis_transformation}
    \rho_{\bf 3}(\gamma) = U_\omega \tilde{\rho}_{\bf 3}(\gamma) U_\omega^\dag\,, \quad
    \tilde{a} = U_\omega^\dag a\,, \quad
    U_\omega = \frac{1}{\sqrt3}\begin{pmatrix}
        1 & 1 & 1 \\ 1 & \omega^2 & \omega \\ 1 & \omega & \omega^2
    \end{pmatrix} \,,
\end{eqnarray}
where $a$ and $\tilde{a}$ are triplets in the complex basis and real basis, respectively.

In the complex basis, the representation matrix satisfies $\rho_{\bf 3}^{\rm T}(\gamma) P_{23} \rho_{\bf 3}(\gamma) = P_{23}$. 
The multiplication rule is listed as follows:
\begin{eqnarray}
\begin{pmatrix}
a_1\\
a_2\\
a_3
\end{pmatrix}_{\bf 3}
\otimes 
\begin{pmatrix}
b_1\\
b_2\\
b_3
\end{pmatrix}_{\bf 3}
&=&\left (a_1b_1+a_2b_3+a_3b_2\right )_{\bf 1} 
\oplus \left (a_3b_3+a_1b_2+a_2b_1\right )_{{\bf 1}'} \nonumber \\
& \oplus& \left (a_2b_2+a_1b_3+a_3b_1\right )_{{\bf 1}''} \nonumber \\
&\oplus& \frac13
\begin{pmatrix}
2a_1b_1-a_2b_3-a_3b_2 \\
2a_3b_3-a_1b_2-a_2b_1 \\
2a_2b_2-a_1b_3-a_3b_1
\end{pmatrix}_{{\bf 3}}
\oplus \frac12
\begin{pmatrix}
a_2b_3-a_3b_2 \\
a_1b_2-a_2b_1 \\
a_3b_1-a_1b_3
\end{pmatrix}_{{\bf 3}} \,, \nonumber \\
a_{{\bf 1}'} \otimes
\begin{pmatrix}
b_1\\
b_2\\
b_3
\end{pmatrix}_{\bf 3} 
&=&
\begin{pmatrix}
a b_3\\
a b_1\\
a b_2
\end{pmatrix}_{\bf 3} \,, \nonumber \\
a_{{\bf 1}''} \otimes
\begin{pmatrix}
b_1\\
b_2\\
b_3
\end{pmatrix}_{\bf 3} 
&=&
\begin{pmatrix}
a b_2\\
a b_3\\
a b_1
\end{pmatrix}_{\bf 3} \ .
\end{eqnarray}
In the real basis, it is given by
\begin{eqnarray}
\begin{pmatrix}
a_1\\
a_2\\
a_3
\end{pmatrix}_{\bf 3}
\otimes 
\begin{pmatrix}
b_1\\
b_2\\
b_3
\end{pmatrix}_{\bf 3}
&=&\left (a_1b_1+a_2b_2+a_3b_3\right )_{\bf 1} 
\oplus \left (a_1b_1+\omega^2 a_2b_2+ \omega a_3b_3\right )_{{\bf 1}'} \nonumber \\
& \oplus& \left (a_1b_1+\omega a_2b_2+\omega^2 a_3b_3\right )_{{\bf 1}''} \nonumber \\
&\oplus& \frac13
\begin{pmatrix}
a_2b_3+a_3b_2 \\
a_3b_1+a_1b_3 \\
a_1b_2+a_2b_1
\end{pmatrix}_{{\bf 3}}
\oplus \frac12
\begin{pmatrix}
a_2b_3-a_3b_2 \\
a_3b_1-a_1b_3 \\
a_1b_2-a_2b_1
\end{pmatrix}_{{\bf 3}\  } \,, \nonumber \\
a_{{\bf 1}'} \otimes
\begin{pmatrix}
b_1\\
b_2\\
b_3
\end{pmatrix}_{\bf 3} 
&=&
\begin{pmatrix}
a b_1\\
\omega a b_2\\
\omega^2 a b_3
\end{pmatrix}_{\bf 3} \,, \nonumber \\
a_{{\bf 1}''} \otimes
\begin{pmatrix}
b_1\\
b_2\\
b_3
\end{pmatrix}_{\bf 3} 
&=&
\begin{pmatrix}
a b_1\\
\omega^2 a b_2\\
\omega a b_3
\end{pmatrix}_{\bf 3}  \ .
\end{eqnarray}

\section{Modular forms of weight 2 for $A_4$}\label{sec:modular-forms}

At $2k=2$, there are three modular forms which form an irreducible triplet of $A_4$. They are expressed in terms of the Dedekind eta functions $\eta(\tau)$ and its derivative:
\begin{eqnarray}
\eta(\tau) = q^{1/24} \prod_{n=1}^{\infty} (1-q^n), q=e^{2 \pi i \tau} \,. 
\end{eqnarray}
In the complex basis, the triplet $Y = (Y_1, Y_2,Y_3)^{\rm T}$ can be expressed as
\begin{eqnarray}
Y_1 (\tau) &=& \frac{i}{2\pi} \left[\frac{\eta'(\frac{\tau}{3})}{\eta(\frac{\tau}{3})} + \frac{\eta'(\frac{\tau+1}{3})}{\eta(\frac{\tau+1}{3})} + \frac{\eta'(\frac{\tau+2}{3})}{\eta(\frac{\tau+3}{3})} - 27\frac{\eta'(3\tau)}{\eta(3\tau)} \right] \,, \nonumber \\
Y_2 (\tau) &=& -\frac{i}{\pi} \left[\frac{\eta'(\frac{\tau}{3})}{\eta(\frac{\tau}{3})} + \omega^2 \frac{\eta'(\frac{\tau+1}{3})}{\eta(\frac{\tau+1}{3})} + \omega \frac{\eta'(\frac{\tau+2}{3})}{\eta(\frac{\tau+3}{3})} \right] \,, \nonumber \\
Y_3 (\tau) &=& -\frac{i}{\pi} \left[\frac{\eta'(\frac{\tau}{3})}{\eta(\frac{\tau}{3})} + \omega \frac{\eta'(\frac{\tau+1}{3})}{\eta(\frac{\tau+1}{3})} + \omega^2 \frac{\eta'(\frac{\tau+2}{3})}{\eta(\frac{\tau+3}{3})} \right] \,. 
\end{eqnarray}
In the real basis, we have $\tilde{Y} = (\tilde{Y}_1, \tilde{Y}_2,\tilde{Y}_3)^{\rm T} = U_\omega^\dag Y $:
\begin{eqnarray}
\tilde{Y}_1 (\tau) &=& \frac{i \sqrt{3} }{2\pi} \left[-\frac{\eta'(\frac{\tau}{3})}{\eta(\frac{\tau}{3})} + \frac{\eta'(\frac{\tau+1}{3})}{\eta(\frac{\tau+1}{3})} + \frac{\eta'(\frac{\tau+2}{3})}{\eta(\frac{\tau+3}{3})} - 9\frac{\eta'(3\tau)}{\eta(3\tau)} \right]\,, \nonumber \\
\tilde{Y}_2 (\tau) &=& \frac{i \sqrt{3} }{2\pi} \left[\frac{\eta'(\frac{\tau}{3})}{\eta(\frac{\tau}{3})} - \frac{\eta'(\frac{\tau+1}{3})}{\eta(\frac{\tau+1}{3})} + \frac{\eta'(\frac{\tau+2}{3})}{\eta(\frac{\tau+3}{3})} - 9\frac{\eta'(3\tau)}{\eta(3\tau)} \right]\,, \nonumber \\
\tilde{Y}_3 (\tau) &=& \frac{i \sqrt{3} }{2\pi} \left[\frac{\eta'(\frac{\tau}{3})}{\eta(\frac{\tau}{3})} + \frac{\eta'(\frac{\tau+1}{3})}{\eta(\frac{\tau+1}{3})} - \frac{\eta'(\frac{\tau+2}{3})}{\eta(\frac{\tau+3}{3})} - 9\frac{\eta'(3\tau)}{\eta(3\tau)} \right] \,. 
\end{eqnarray}

\section{Vacuum alignments for $\Phi$}
\label{sec:vacuum-alignments}

In this appendix, we discuss how to break $A_4^l \times A_4^\nu$ to a single $A_4$ via the VEV of a bi-triplet $\Phi$. 

Vacuum alignments for the bi-triplet scalar $\Phi$ can be realised using the driving field method. It has been proved in \cite{deMedeirosVarzielas:2019cyj} that in the framework of $S_4 \times S_4$, by introducing one bi-triplet and one triplet driving fields, the minimisation of the superpotential gives 24 solutions, each of them mapping to an element of $S_4$. All these VEVs are equivalent to each other following a group transformation of $S_4$ and lead to the breaking of $S_4 \times S_4$ to a single $S_4$. Thus, one can fix the VEV $\langle \Phi \rangle \propto P_{23}$ without loss of generality. The situation in $A_4 \times A_4$ is slightly different and we will discuss below.

We consider the same setup as used in \cite{deMedeirosVarzielas:2021pug}, one bi-triplet driving field $\chi_{l\nu} \sim ({\bf 3}, {\bf 3})$ and one triplet driving field $\chi_\nu \sim ({\bf 1},{\bf 3})$ (or alternatively, $\chi_l \sim ({\bf 3},{\bf 1})$) of $A_4^l \times A_4^\nu$. A renormalisable superpotential terms for the vacuum alignment are given by
\begin{eqnarray}
w_d &=& \left[g_S (\Phi\Phi)_{({\bf 3}_S, {\bf 3}_S)} + g_A (\Phi\Phi)_{({\bf 3}_A, {\bf 3}_A)} + {\rm M} \Phi \right] \chi_{l\nu} + g_S' (\Phi\Phi)_{({\bf 1},{\bf 3})} \chi_{\nu} \,, 
\end{eqnarray}
where ${\rm M}$ is a  mass-dimensional coefficient, $g_S$ and $g_A$ are dimensionless parameters. 
Minimisation of the superpotential gives rise to equations, in the real triplet basis, as
\begin{eqnarray}
&&\sum_{j,k=1,2,3;} \sum_{\beta,\gamma=1,2,3} (g_S |\epsilon_{\alpha\beta\gamma}| |\epsilon_{ijk}| + 
g_A \epsilon_{\alpha\beta\gamma} \epsilon_{ijk}) \tilde{\Phi}_{\beta j} \tilde{\Phi}_{\gamma k}
+ {\rm M} \tilde{\Phi}_{\alpha i} = 0\,,\nonumber\\
&&\sum_{j,k=1,2,3;} \sum_{\beta=1,2,3} |\epsilon_{ijk}| \tilde{\Phi}_{\beta j} \tilde{\Phi}_{\beta k} = 0
\end{eqnarray}
for $\alpha=1,2,3$ and $i=1,2,3$. For general values of $g_S$ and $g_A$, with $g_S \neq g_A$ and $g_S \neq -g_A$, these equations gives 24 non-trivial solutions, instead of 12. As summarised in \cite{deMedeirosVarzielas:2021pug}, they are classified into two sets, 
\begin{eqnarray} \label{eq:vacuum_solutions}
\langle \tilde{\Phi} \rangle & \in &\left\{
\left(
\begin{array}{ccc}
 1 & 0 & 0 \\
 0 & 1 & 0 \\
 0 & 0 & 1 \\
\end{array}
\right),\left(
\begin{array}{ccc}
 1 & 0 & 0 \\
 0 & -1 & 0 \\
 0 & 0 & -1 \\
\end{array}
\right),\left(
\begin{array}{ccc}
 -1 & 0 & 0 \\
 0 & -1 & 0 \\
 0 & 0 & 1 \\
\end{array}
\right),\left(
\begin{array}{ccc}
 -1 & 0 & 0 \\
 0 & 1 & 0 \\
 0 & 0 & -1 \\
\end{array}
\right),\right. \nonumber\\ &&
\left(
\begin{array}{ccc}
 0 & 0 & 1 \\
 1 & 0 & 0 \\
 0 & 1 & 0 \\
\end{array}
\right),\left(
\begin{array}{ccc}
 0 & 0 & -1 \\
 -1 & 0 & 0 \\
 0 & 1 & 0 \\
\end{array}
\right),\left(
\begin{array}{ccc}
 0 & 0 & -1 \\
 1 & 0 & 0 \\
 0 & -1 & 0 \\
\end{array}
\right),\left(
\begin{array}{ccc}
 0 & 0 & 1 \\
 -1 & 0 & 0 \\
 0 & -1 & 0 \\
\end{array}
\right),\nonumber\\ &&
\left.\left(
\begin{array}{ccc}
 0 & 1 & 0 \\
 0 & 0 & 1 \\
 1 & 0 & 0 \\
\end{array}
\right),\left(
\begin{array}{ccc}
 0 & 1 & 0 \\
 0 & 0 & -1 \\
 -1 & 0 & 0 \\
\end{array}
\right),\left(
\begin{array}{ccc}
 0 & -1 & 0 \\
 0 & 0 & 1 \\
 -1 & 0 & 0 \\
\end{array}
\right),\left(
\begin{array}{ccc}
 0 & -1 & 0 \\
 0 & 0 & -1 \\
 1 & 0 & 0 \\
\end{array}
\right)\right\} v_{\Phi}\nonumber\\ 
&\cup&
\left\{\left(
\begin{array}{ccc}
 1 & 0 & 0 \\
 0 & 0 & 1 \\
 0 & 1 & 0 \\
\end{array}
\right),\left(
\begin{array}{ccc}
 1 & 0 & 0 \\
 0 & 0 & -1 \\
 0 & -1 & 0 \\
\end{array}
\right),\left(
\begin{array}{ccc}
 -1 & 0 & 0 \\
 0 & 0 & 1 \\
 0 & -1 & 0 \\
\end{array}
\right),\left(
\begin{array}{ccc}
 -1 & 0 & 0 \\
 0 & 0 & -1 \\
 0 & 1 & 0 \\
\end{array}
\right),\right.\nonumber\\ &&
\left(
\begin{array}{ccc}
 0 & 1 & 0 \\
 1 & 0 & 0 \\
 0 & 0 & 1 \\
\end{array}
\right),\left(
\begin{array}{ccc}
 0 & 1 & 0 \\
 -1 & 0 & 0 \\
 0 & 0 & -1 \\
\end{array}
\right),\left(
\begin{array}{ccc}
 0 & -1 & 0 \\
 1 & 0 & 0 \\
 0 & 0 & -1 \\
\end{array}
\right),\left(
\begin{array}{ccc}
 0 & -1 & 0 \\
 -1 & 0 & 0 \\
 0 & 0 & 1 \\
\end{array}
\right),\nonumber\\ &&
\left.\left(
\begin{array}{ccc}
 0 & 0 & 1 \\
 0 & 1 & 0 \\
 1 & 0 & 0 \\
\end{array}
\right),\left(
\begin{array}{ccc}
 0 & 0 & -1 \\
 0 & -1 & 0 \\
 1 & 0 & 0 \\
\end{array}
\right),\left(
\begin{array}{ccc}
 0 & 0 & -1 \\
 0 & 1 & 0 \\
 -1 & 0 & 0 \\
\end{array}
\right),\left(
\begin{array}{ccc}
 0 & 0 & 1 \\
 0 & -1 & 0 \\
 -1 & 0 & 0 \\
\end{array}
\right)\right\} v_{\Phi}\,,
\end{eqnarray}
where $v_{\Phi}$ is a constant which satisfies $v_{\Phi} \propto {\rm M}/(g_S + g_A)$ for the first set and $v_{\Phi} \propto {\rm M}/(g_S - g_A)$ for the second set. In the specific case $g_S = g_A$, only solutions of the first set are valid. Similarly, if $g_S = -g_A$, only solutions of the second set are valid. 

The first set correspond to all 12 elements of $A_4$ in the triplet real basis, i.e., $\tilde{\rho}_{\bf 3}(\gamma) v_\Phi$. The second set are different from  elements of $A_4$ by an permutation matrix, e.g., $\tilde{\rho}_{\bf 3}(\gamma) P_{23} v_\Phi$. 

Using the basis transformation, we obtain solutions for $\langle \Phi \rangle_{i\alpha} = \sum_{j,\beta} (U_\omega)_{ij} (U_\omega)_{\alpha\beta} \langle \tilde{\Phi} \rangle_{j\beta}$ in the complex basis. With the help of $U_\omega^\dag = U_\omega P_{23}$, we can express all these solutions as 
\begin{eqnarray}
\langle \Phi \rangle = \rho_{\mathbf{3}}(\gamma) P_{23} v_{\Phi}, \rho_{\mathbf{3}}(\gamma) v_{\Phi}
\end{eqnarray}
with $\gamma$ being any element of $A_4$. Below, we prove that any of these VEVs breaks two $A_4$'s to a single $A_4$. 

Given any modular transformation $\gamma_l \in A_4^l$ and $\gamma_\nu \in A_4^\nu$, the Dirac Yukawa coupling term for neutrinos $(\nu^c \Phi L)_{({\bf 1,1})} H_u = (\nu^c)^{\rm T} P_{23} \Phi P_{23} L H_u$, before $\Phi$ gains the VEV, transforms as 
\begin{eqnarray}
(\nu^c)^{\rm T} P_{23} \Phi P_{23} L H_u \to (\nu^c)^{\rm T} \rho_{\bf 3}^{\rm T}(\gamma_\nu) P_{23} \rho_{\bf 3}(\gamma_\nu) \Phi \rho_{\bf 3}^{\rm T} (\gamma_l) P_{23} \rho_{\bf 3}(\gamma_l) L H_u \,.
\end{eqnarray}
Using the identity $\rho_{\bf 3}^{\rm T} P_{23} \rho_{\bf 3} = P_{23}$, we see that
the right-hand side is identical to the left-hand side.
Now let us fix $\Phi$ at one of VEVs and discuss the breaking of two $A_4$'s. 
\begin{itemize}
\item If $\Phi$ takes a VEV of the first set, i.e., $\rho_{\mathbf{3}}(\gamma) P_{23} v_{\Phi}$, the coupling becomes  $(\nu^c)^{\rm T} P_{23} \rho_{\bf 3}(\gamma) L v_\Phi H_u$. Under the modular transformation, we have
\begin{eqnarray} \label{eq:modular_1st}
(\nu^c)^{\rm T} P_{23} \rho_{\bf 3}(\gamma) L \to (\nu^c)^{\rm T} \rho_{\bf 3}^{\rm T}(\gamma_\nu) P_{23} \rho_{\bf 3}(\gamma) \rho_{\bf 3}(\gamma_l) L = 
(\nu^c)^{\rm T} \rho_{\bf 3}^{\rm T}(\gamma_\nu) P_{23} \rho_{\bf 3}(\gamma \gamma_l \gamma^{-1}) \rho_{\bf 3}(\gamma) L \,.
\end{eqnarray}
The right-hand side is identical to the left-hand side only if $\gamma_\nu = \gamma \gamma_l \gamma^{-1}$. Namely, the modular transformation in the neutrino sector is conjugate to that in the lepton sector. It is convenient to re-write $\nu^c$ in a new basis $\nu^c_\gamma = \rho_{\bf 3}^{-1}(\gamma) \nu^c$, and then $(\nu^c)^{\rm T} P_{23} \rho_{\bf 3}(\gamma) L  = (\nu_\gamma^c)^{\rm T} P_{23} L$.  Once the conjugacy relation between $\gamma_\nu$ and $\gamma_l$ is imposed, a modular transformation in the lepton sector $L \to \rho_{\bf 3}(\gamma_l) L$ results in a modular transformation in the neutrino sector $\nu_\gamma^c \to \rho_{\bf 3}(\gamma_l) \nu_\gamma^c$, and $(\nu_\gamma^c)^{\rm T} P_{23} L$ is invariant. Here and below, we ignore flavour-independent overall factors $(c\tau + d)^{2k}$. The Majorana mass term $\frac{1}{2} (\nu^c)^{\rm T} M_{\nu^c}(\tau_\nu) \nu^c$  after the basis transformation is given by
\begin{eqnarray}
     \frac{1}{2} (\nu^c_\gamma)^{\rm T} M_{\nu^c}(\gamma\tau_\nu) \nu^c_\gamma \,.
\end{eqnarray}
Therefore, a VEV at $\rho_{\bf 3}(\gamma) P_{23} v_\Phi$ is equivalent to fixing the VEV at $P_{23} v_\Phi$ and shifting the Majorana mass from $M_{\nu^c}(\tau_\nu)$ to $M_{\nu^c}(\gamma\tau_\nu)$ after the basis of right-handed neutrino transforming from $\nu^c$ to $\nu^c_{\gamma} = \rho_{\bf 3}^{-1} (\gamma) \nu^c$. 

\item If $\Phi$ takes a VEV of the second set, i.e., $\rho_{\mathbf{3}}(\gamma) v_{\Phi}$, the coupling becomes  $(\nu^c)^{\rm T} P_{23} \rho_{\bf 3}(\gamma) P_{23} L v_\Phi H_u$. Under the modular transformation, we have
\begin{eqnarray} \label{eq:modular_2nd}
(\nu^c)^{\rm T} P_{23} \rho_{\bf 3}(\gamma) P_{23} L \to (\nu^c)^{\rm T} \rho_{\bf 3}^{\rm T}(\gamma_\nu) P_{23} \rho_{\bf 3}(\gamma) P_{23} \rho_{\bf 3}(\gamma_l) L \,.
\end{eqnarray}
Here $P_{23}$ is not a representation matrix of any element of $A_4$, but a representation matrix of an outer automorphism. We denote it as $\rho_{\bf 3}(U)$. The automorphism group of $A_4$ satisfies ${\rm Aut}(A_4) \simeq S_4$, the inner automorphism group ${\rm Inn} (A_4) \simeq A_4$, the outer automorphism group ${\rm Out}(A_4) \equiv {\rm Aut}(A_4)/ {\rm Inn}(A_4) \simeq Z_2 = \{1, U \}$, and $U$ can be regarded as the third generator of $S_4$ which satisfies $U^2 = (SU)^2 = (TU)^2 = (STU)^4 = 1$. Therefore, for any $\gamma_l$ of $A_4$, there will always be an element $\gamma'_l = U \gamma_l U$ of $A_4$. The right-hand side of the above formula equals
\begin{eqnarray}
(\nu^c)^{\rm T} \rho_{\bf 3}^{\rm T}(\gamma_\nu) P_{23} \rho_{\bf 3}(\gamma) \rho_{\bf 3}(\gamma_l') P_{23} L  = 
(\nu^c)^{\rm T} \rho_{\bf 3}^{\rm T}(\gamma_\nu) P_{23} \rho_{\bf 3}(\gamma \gamma_l' \gamma^{-1}) \rho_{\bf 3}(\gamma) P_{23} L \,.
\end{eqnarray}
It is identical to the left-hand side of Eq.~\eqref{eq:modular_2nd} only if $\gamma_\nu = \gamma \gamma_l' \gamma^{-1} = (\gamma U) \gamma_l (\gamma U)^{-1}$. For any modular transformation in the charged lepton sector, there is always a follow-up modular transformation in the neutrino sector to keep the Yukawa coupling invariant under a single $A_4$ symmetry. However, due to the involvement of $U$, the modular transformation in the neutrino sector is not conjugate to that in the lepton sector. 
Using $P_{23} \rho_{\bf 3}(\gamma) =\rho_{\bf 3}^{\rm T}(\gamma^{-1})  P_{23}$ and considering the basis transformation 
$\nu^{\prime c}_{\gamma} = P_{23}\rho_{\bf 3}^{-1} (\gamma) \nu^c$, we may re-write
$(\nu^c)^{\rm T} P_{23} \rho_{\bf 3}(\gamma) P_{23} L$ in a much simpler form 
$(\nu^{\prime c}_\gamma)^{\rm T} P_{23} L$. Note that in this case, it is $\nu_\gamma^{\prime c}$, not $\nu_\gamma^c$, following the same modular transformation as that in the charged lepton sector. In the new basis, the Majorana mass term becomes
\begin{eqnarray}
    \frac{1}{2} (\nu^{\prime c}_\gamma)^{\rm T} M_{\nu^c}(\gamma\tau_\nu) \nu^{\prime c}_\gamma \,.
\end{eqnarray}
Therefore, we have proven that a VEV at $\rho_{\rm 3} (\gamma) v_\Phi$ is equivalent to fixing the VEV at $P_{23} v_\Phi$ and shifting Majorana mass from $M_{\nu^c}(\tau_\nu)$ to $M_{\nu^c}(\gamma\tau_\nu)$ after the basis of right-handed neutrino transforming from $\nu^c$ to $\nu^{\prime c}_{\gamma} = P_{23}\rho_{\bf 3}^{-1} (\gamma) \nu^c$. 

\end{itemize}


\begin{thebibliography}{99}

\bibitem{Altarelli:2005yp}
G.~Altarelli and F.~Feruglio,
Nucl. Phys. B \textbf{720} (2005), 64-88
doi:10.1016/j.nuclphysb.2005.05.005
[arXiv:hep-ph/0504165 [hep-ph]].

\bibitem{Altarelli:2005yx}
G.~Altarelli and F.~Feruglio,
Nucl. Phys. B \textbf{741} (2006), 215-235
doi:10.1016/j.nuclphysb.2006.02.015
[arXiv:hep-ph/0512103 [hep-ph]].

\bibitem{Ge:2011ih}
S.~F.~Ge, D.~A.~Dicus and W.~W.~Repko,
Phys. Lett. B \textbf{702} (2011), 220-223
doi:10.1016/j.physletb.2011.06.096
[arXiv:1104.0602 [hep-ph]].

\bibitem{Ge:2011qn}
S.~F.~Ge, D.~A.~Dicus and W.~W.~Repko,
Phys. Rev. Lett. \textbf{108} (2012), 041801
doi:10.1103/PhysRevLett.108.041801
[arXiv:1108.0964 [hep-ph]].

\bibitem{Albright:2008rp}
C.~H.~Albright and W.~Rodejohann,
Eur. Phys. J. C \textbf{62} (2009), 599-608
doi:10.1140/epjc/s10052-009-1074-3
[arXiv:0812.0436 [hep-ph]].

\bibitem{Grimus:2008tt}
W.~Grimus and L.~Lavoura,
JHEP \textbf{09} (2008), 106
doi:10.1088/1126-6708/2008/09/106
[arXiv:0809.0226 [hep-ph]].

\bibitem{Harrison:2002er}
P.~F.~Harrison, D.~H.~Perkins and W.~G.~Scott,
Phys. Lett. B \textbf{530} (2002), 167
doi:10.1016/S0370-2693(02)01336-9
[arXiv:hep-ph/0202074 [hep-ph]].

\bibitem{Xing:2002sw}
Z.~z.~Xing,
Phys. Lett. B \textbf{533} (2002), 85-93
doi:10.1016/S0370-2693(02)01649-0
[arXiv:hep-ph/0204049 [hep-ph]].

\bibitem{Feruglio:2017spp}
F.~Feruglio,
doi:10.1142/9789813238053\_0012
[arXiv:1706.08749 [hep-ph]].

\bibitem{Kobayashi:2018vbk}
T.~Kobayashi, K.~Tanaka and T.~H.~Tatsuishi,
Phys. Rev. D \textbf{98} (2018) no.1, 016004
doi:10.1103/PhysRevD.98.016004
[arXiv:1803.10391 [hep-ph]].

\bibitem{Criado:2018thu}
J.~C.~Criado and F.~Feruglio,
SciPost Phys. \textbf{5} (2018) no.5, 042
doi:10.21468/SciPostPhys.5.5.042
[arXiv:1807.01125 [hep-ph]].

\bibitem{Penedo:2018nmg}
J.~T.~Penedo and S.~T.~Petcov,
Nucl. Phys. B \textbf{939} (2019), 292-307
doi:10.1016/j.nuclphysb.2018.12.016
[arXiv:1806.11040 [hep-ph]].

\bibitem{Novichkov:2018nkm}
P.~P.~Novichkov, J.~T.~Penedo, S.~T.~Petcov and A.~V.~Titov,
JHEP \textbf{04} (2019), 174
doi:10.1007/JHEP04(2019)174
[arXiv:1812.02158 [hep-ph]].

\bibitem{Ding:2020msi}
G.~J.~Ding, S.~F.~King, C.~C.~Li and Y.~L.~Zhou,
JHEP \textbf{08} (2020), 164
doi:10.1007/JHEP08(2020)164
[arXiv:2004.12662 [hep-ph]].

\bibitem{Kobayashi:2018wkl}
T.~Kobayashi, Y.~Shimizu, K.~Takagi, M.~Tanimoto, T.~H.~Tatsuishi and H.~Uchida,
Phys. Lett. B \textbf{794} (2019), 114-121
doi:10.1016/j.physletb.2019.05.034
[arXiv:1812.11072 [hep-ph]].

\bibitem{Kobayashi:2019rzp}
T.~Kobayashi, Y.~Shimizu, K.~Takagi, M.~Tanimoto and T.~H.~Tatsuishi,
PTEP \textbf{2020} (2020) no.5, 053B05
doi:10.1093/ptep/ptaa055
[arXiv:1906.10341 [hep-ph]].

\bibitem{Okada:2019xqk}
H.~Okada and Y.~Orikasa,
Phys. Rev. D \textbf{100} (2019) no.11, 115037
doi:10.1103/PhysRevD.100.115037
[arXiv:1907.04716 [hep-ph]].

\bibitem{Kobayashi:2018scp}
T.~Kobayashi, N.~Omoto, Y.~Shimizu, K.~Takagi, M.~Tanimoto and T.~H.~Tatsuishi,
JHEP \textbf{11} (2018), 196
doi:10.1007/JHEP11(2018)196
[arXiv:1808.03012 [hep-ph]].

\bibitem{deAnda:2018ecu}
F.~J.~de Anda, S.~F.~King and E.~Perdomo,
Phys. Rev. D \textbf{101} (2020) no.1, 015028
doi:10.1103/PhysRevD.101.015028
[arXiv:1812.05620 [hep-ph]].

\bibitem{Okada:2018yrn}
H.~Okada and M.~Tanimoto,
Phys. Lett. B \textbf{791} (2019), 54-61
doi:10.1016/j.physletb.2019.02.028
[arXiv:1812.09677 [hep-ph]].

\bibitem{Novichkov:2018yse}
P.~P.~Novichkov, S.~T.~Petcov and M.~Tanimoto,
Phys. Lett. B \textbf{793} (2019), 247-258
doi:10.1016/j.physletb.2019.04.043
[arXiv:1812.11289 [hep-ph]].

\bibitem{Nomura:2019jxj}
T.~Nomura and H.~Okada,
Phys. Lett. B \textbf{797} (2019), 134799
doi:10.1016/j.physletb.2019.134799
[arXiv:1904.03937 [hep-ph]].

\bibitem{Okada:2019uoy}
H.~Okada and M.~Tanimoto,
Eur. Phys. J. C \textbf{81} (2021) no.1, 52
doi:10.1140/epjc/s10052-021-08845-y
[arXiv:1905.13421 [hep-ph]].

\bibitem{Nomura:2019yft}
T.~Nomura and H.~Okada,
Nucl. Phys. B \textbf{966} (2021), 115372
doi:10.1016/j.nuclphysb.2021.115372
[arXiv:1906.03927 [hep-ph]].

\bibitem{Ding:2019zxk}
G.~J.~Ding, S.~F.~King and X.~G.~Liu,
JHEP \textbf{09} (2019), 074
doi:10.1007/JHEP09(2019)074
[arXiv:1907.11714 [hep-ph]].

\bibitem{Okada:2019mjf}
H.~Okada and Y.~Orikasa,
[arXiv:1907.13520 [hep-ph]].

\bibitem{Nomura:2019lnr}
T.~Nomura, H.~Okada and O.~Popov,
Phys. Lett. B \textbf{803} (2020), 135294
doi:10.1016/j.physletb.2020.135294
[arXiv:1908.07457 [hep-ph]].

\bibitem{Kobayashi:2019xvz}
T.~Kobayashi, Y.~Shimizu, K.~Takagi, M.~Tanimoto and T.~H.~Tatsuishi,
Phys. Rev. D \textbf{100} (2019) no.11, 115045
[erratum: Phys. Rev. D \textbf{101} (2020) no.3, 039904]
doi:10.1103/PhysRevD.100.115045
[arXiv:1909.05139 [hep-ph]].

\bibitem{Asaka:2019vev}
T.~Asaka, Y.~Heo, T.~H.~Tatsuishi and T.~Yoshida,
JHEP \textbf{01} (2020), 144
doi:10.1007/JHEP01(2020)144
[arXiv:1909.06520 [hep-ph]].

\bibitem{Ding:2019gof}
G.~J.~Ding, S.~F.~King, X.~G.~Liu and J.~N.~Lu,
JHEP \textbf{12} (2019), 030
doi:10.1007/JHEP12(2019)030
[arXiv:1910.03460 [hep-ph]].

\bibitem{Zhang:2019ngf}
D.~Zhang,
Nucl. Phys. B \textbf{952} (2020), 114935
doi:10.1016/j.nuclphysb.2020.114935
[arXiv:1910.07869 [hep-ph]].

\bibitem{Nomura:2019xsb}
T.~Nomura, H.~Okada and S.~Patra,
Nucl. Phys. B \textbf{967} (2021), 115395
doi:10.1016/j.nuclphysb.2021.115395
[arXiv:1912.00379 [hep-ph]].

\bibitem{Wang:2019xbo}
X.~Wang,
Nucl. Phys. B \textbf{957} (2020), 115105
doi:10.1016/j.nuclphysb.2020.115105
[arXiv:1912.13284 [hep-ph]].

\bibitem{Kobayashi:2019gtp}
T.~Kobayashi, T.~Nomura and T.~Shimomura,
Phys. Rev. D \textbf{102} (2020) no.3, 035019
doi:10.1103/PhysRevD.102.035019
[arXiv:1912.00637 [hep-ph]].

\bibitem{King:2020qaj}
S.~J.~D.~King and S.~F.~King,
JHEP \textbf{09} (2020), 043
doi:10.1007/JHEP09(2020)043
[arXiv:2002.00969 [hep-ph]].

\bibitem{Ding:2020yen}
G.~J.~Ding and F.~Feruglio,
JHEP \textbf{06} (2020), 134
doi:10.1007/JHEP06(2020)134
[arXiv:2003.13448 [hep-ph]].

\bibitem{Okada:2020rjb}
H.~Okada and M.~Tanimoto,
Phys. Dark Univ. \textbf{40} (2023), 101204
doi:10.1016/j.dark.2023.101204
[arXiv:2005.00775 [hep-ph]].

\bibitem{Nomura:2020opk}
T.~Nomura and H.~Okada,
JCAP \textbf{09} (2022), 049
doi:10.1088/1475-7516/2022/09/049
[arXiv:2007.04801 [hep-ph]].

\bibitem{Asaka:2020tmo}
T.~Asaka, Y.~Heo and T.~Yoshida,
Phys. Lett. B \textbf{811} (2020), 135956
doi:10.1016/j.physletb.2020.135956
[arXiv:2009.12120 [hep-ph]].

\bibitem{Okada:2020brs}
H.~Okada and M.~Tanimoto,
JHEP \textbf{03} (2021), 010
doi:10.1007/JHEP03(2021)010
[arXiv:2012.01688 [hep-ph]].

\bibitem{Yao:2020qyy}
C.~Y.~Yao, J.~N.~Lu and G.~J.~Ding,
JHEP \textbf{05} (2021), 102
doi:10.1007/JHEP05(2021)102
[arXiv:2012.13390 [hep-ph]].

\bibitem{Feruglio:2021dte}
F.~Feruglio, V.~Gherardi, A.~Romanino and A.~Titov,
JHEP \textbf{05} (2021), 242
doi:10.1007/JHEP05(2021)242
[arXiv:2101.08718 [hep-ph]].

\bibitem{Novichkov:2018ovf}
P.~P.~Novichkov, J.~T.~Penedo, S.~T.~Petcov and A.~V.~Titov,
JHEP \textbf{04} (2019), 005
doi:10.1007/JHEP04(2019)005
[arXiv:1811.04933 [hep-ph]].

\bibitem{deMedeirosVarzielas:2019cyj}
I.~de Medeiros Varzielas, S.~F.~King and Y.~L.~Zhou,
Phys. Rev. D \textbf{101} (2020) no.5, 055033
doi:10.1103/PhysRevD.101.055033
[arXiv:1906.02208 [hep-ph]].

\bibitem{Kobayashi:2019mna}
T.~Kobayashi, Y.~Shimizu, K.~Takagi, M.~Tanimoto and T.~H.~Tatsuishi,
JHEP \textbf{02} (2020), 097
doi:10.1007/JHEP02(2020)097
[arXiv:1907.09141 [hep-ph]].

\bibitem{Criado:2019tzk}
J.~C.~Criado, F.~Feruglio and S.~J.~D.~King,
JHEP \textbf{02} (2020), 001
doi:10.1007/JHEP02(2020)001
[arXiv:1908.11867 [hep-ph]].

\bibitem{King:2019vhv}
S.~F.~King and Y.~L.~Zhou,
Phys. Rev. D \textbf{101} (2020) no.1, 015001
doi:10.1103/PhysRevD.101.015001
[arXiv:1908.02770 [hep-ph]].

\bibitem{Wang:2019ovr}
X.~Wang and S.~Zhou,
JHEP \textbf{05} (2020), 017
doi:10.1007/JHEP05(2020)017
[arXiv:1910.09473 [hep-ph]].

\bibitem{Wang:2020dbp}
X.~Wang,
Nucl. Phys. B \textbf{962} (2021), 115247
doi:10.1016/j.nuclphysb.2020.115247
[arXiv:2007.05913 [hep-ph]].

\bibitem{Qu:2021jdy}
B.~Y.~Qu, X.~G.~Liu, P.~T.~Chen and G.~J.~Ding,
Phys. Rev. D \textbf{104} (2021) no.7, 076001
doi:10.1103/PhysRevD.104.076001
[arXiv:2106.11659 [hep-ph]].

\bibitem{Ding:2019xna}
G.~J.~Ding, S.~F.~King and X.~G.~Liu,
Phys. Rev. D \textbf{100} (2019) no.11, 115005
doi:10.1103/PhysRevD.100.115005
[arXiv:1903.12588 [hep-ph]].

\bibitem{Ding:2023htn}
G.~J.~Ding and S.~F.~King,
[arXiv:2311.09282 [hep-ph]].

\bibitem{deMedeirosVarzielas:2021pug}
I.~de Medeiros Varzielas and J.~Louren\c{c}o,
Nucl. Phys. B \textbf{979} (2022), 115793
doi:10.1016/j.nuclphysb.2022.115793
[arXiv:2107.04042 [hep-ph]].

\bibitem{Kikuchi:2023jap}
S.~Kikuchi, T.~Kobayashi, K.~Nasu, S.~Takada and H.~Uchida,
JHEP \textbf{07} (2023), 134
doi:10.1007/JHEP07(2023)134
[arXiv:2302.03326 [hep-ph]].

\bibitem{Petcov:2023vws}
S.~T.~Petcov and M.~Tanimoto,
JHEP \textbf{08} (2023), 086
doi:10.1007/JHEP08(2023)086
[arXiv:2306.05730 [hep-ph]].

\bibitem{deMedeirosVarzielas:2023crv}
I.~de Medeiros Varzielas, M.~Levy, J.~T.~Penedo and S.~T.~Petcov,
JHEP \textbf{09} (2023), 196
doi:10.1007/JHEP09(2023)196
[arXiv:2307.14410 [hep-ph]].

\bibitem{King:2021fhl}
S.~F.~King and Y.~L.~Zhou,
JHEP \textbf{04} (2021), 291
doi:10.1007/JHEP04(2021)291
[arXiv:2103.02633 [hep-ph]].

\bibitem{deMedeirosVarzielas:2023ujt}
I.~de Medeiros Varzielas, S.~F.~King and M.~Levy,
[arXiv:2309.15901 [hep-ph]].

\bibitem{King:2023snq}
S.~F.~King and X.~Wang,
[arXiv:2310.10369 [hep-ph]].

\bibitem{Chen:2019ewa}
M.~C.~Chen, S.~Ramos-S\'anchez and M.~Ratz,
Phys. Lett. B \textbf{801} (2020), 135153
doi:10.1016/j.physletb.2019.135153
[arXiv:1909.06910 [hep-ph]].

\bibitem{Ferrara:1989bc}
S.~Ferrara, D.~Lust, A.~D.~Shapere and S.~Theisen,
Phys. Lett. B \textbf{225} (1989), 363
doi:10.1016/0370-2693(89)90583-2

\bibitem{deMedeirosVarzielas:2020kji}
I.~de Medeiros Varzielas, M.~Levy and Y.~L.~Zhou,
JHEP \textbf{11} (2020), 085
doi:10.1007/JHEP11(2020)085
[arXiv:2008.05329 [hep-ph]].

\bibitem{Esteban:2020cvm}
I.~Esteban, M.~C.~Gonzalez-Garcia, M.~Maltoni, T.~Schwetz and A.~Zhou,
JHEP \textbf{09} (2020), 178
doi:10.1007/JHEP09(2020)178
[arXiv:2007.14792 [hep-ph]].

\bibitem{web_link}
NuFIT 5.2 (2022), www.nu-fit.org.

\bibitem{Xing:2008fg}
Z.~z.~Xing and S.~Zhou,
Phys. Lett. B \textbf{666} (2008), 166-172
doi:10.1016/j.physletb.2008.07.011
[arXiv:0804.3512 [hep-ph]].

\bibitem{Xing:2014zka}
Z.~z.~Xing and S.~Zhou,
Phys. Lett. B \textbf{737} (2014), 196-200
doi:10.1016/j.physletb.2014.08.047
[arXiv:1404.7021 [hep-ph]].

\bibitem{Xing:2015zha}
Z.~z.~Xing, Z.~h.~Zhao and Y.~L.~Zhou,
Eur. Phys. J. C \textbf{75} (2015) no.9, 423
doi:10.1140/epjc/s10052-015-3656-6
[arXiv:1504.05820 [hep-ph]].

\bibitem{Cao:2019hli}
J.~Cao, G.~Y.~Huang, Y.~F.~Li, Y.~Wang, L.~J.~Wen, Z.~Z.~Xing, Z.~H.~Zhao and S.~Zhou,
Chin. Phys. C \textbf{44} (2020) no.3, 031001
doi:10.1088/1674-1137/44/3/031001
[arXiv:1908.08355 [hep-ph]].

\bibitem{Denton:2023hkx}
P.~B.~Denton and J.~Gehrlein,
[arXiv:2308.09737 [hep-ph]].

\end{thebibliography}
\end{document}